\documentclass[aps,prl,twocolumn,amssymb,superscriptaddress]{revtex4-2}
\usepackage{subfigure}
\usepackage{graphicx}
\usepackage{sidecap}
\usepackage{soul}
\usepackage{color}
\usepackage[usenames,dvipsnames]{xcolor}
\usepackage{rotating}
 \usepackage{amsmath,amsthm}
 \usepackage{enumitem}
\usepackage{epstopdf}
\usepackage{cancel}
\DeclareMathOperator*{\argmax}{argmax} 

\begin{document}
\title{Emergence of metachronal waves in a chain of symmetrically beating filaments}
\author{Narina Jung}
\thanks{njung@kias.re.kr}
\affiliation{Korea Institute for Advanced Study, Seoul 02455, Korea}
\author{Won Kyu Kim}
\affiliation{Korea Institute for Advanced Study, Seoul 02455, Korea}
\author{Changbong Hyeon}
\thanks{hyeoncb@kias.re.kr}
\affiliation{Korea Institute for Advanced Study, Seoul 02455, Korea}
\date{\today}

\begin{abstract}
Recent experiments have shown that metachronal waves (MCWs) can emerge from a chain of symmetrically beating nematodes aligned at the edge of sessile droplets. 
Our study, employing a coupled elastohydrodynamic model of active filaments, elucidates that a misalignment caused by a tilt against the bounding wall disrupts the synchronization and generates a constant time lag between adjacent filaments, giving rise to MCWs. 
The MCWs, enhancing the fluid circulation, achieve their maximum thermodynamic efficiency over the same range of tilt angles observed in the nematode experiments. 
\end{abstract}
\maketitle

{\it Introduction.} 
Collective dynamics observed in animal population are an outcome of local dynamics and coupling between individual agents 
\cite{couzin2007collective,elgeti2015physics,cavagna2014bird,moussaid2011simple,lee2018PRE,hong2020JSM}. 
Among them, long wavelength modulations of collective beating behavior called \emph{metachronal waves} (MCWs), characterized by a constant phase or time lag between the dynamics of adjacent agents,   
have long been a topic of abiding interest~\cite{taylor1951analysis, taylor1952analysis,gueron1997cilia,gueron1999energetic,niedermayer2008synchronization,gu2020magnetic,wang2024programmable}. 

Since the Taylor's seminal works~\cite{taylor1951analysis, taylor1952analysis}, 
hydrodynamic interactions (HIs) have widely been appreciated as the key for the coordination and generation of MCWs among filamentous micro-swimmers in low Reynolds number environment~\cite{hancock1953self,yang2008cooperation,golestanian2011hydrodynamic,kim2006pumping,gueron1997cilia,gueron1999energetic,zheng2023self}. 
Significance of short-range steric interactions (SIs) was also highlighted in the gait synchronization and generation of collective waves
among \emph{C. elegans}~\cite{yuan2014gait,chelakkot2021synchronized}. 
Applying transverse external flow to symmetrically undulating filaments, 
Guirao and Joanny have suggested that breaking the left-right (LR) symmetry in the filament dynamics as well as including the HIs is necessary to generate MCWs~\cite{guirao2007spontaneous}. 

In broadly discussed MCWs in cilia, individual cilium exhibits two-phase asymmetric beating consisting of power and recovery strokes.   
However, 
MCWs that Peshkov \emph{et al.} have recently demonstrated in a chain of \emph{Turbatrix aceti} (\emph{T. aceti})~\cite{peshkov2022synchronized,quillen2021metachronal} 
are puzzling in that the nematodes exhibit symmetric beating. 
In this Letter, 
we extend a phenomenological elastohydrodynamic model of a   filament~\cite{goldstein2016elastohydrodynamic} to many filaments, and study the physics underlying the formation of MCWs and its energetics associated with the flow of surrounding fluid.

\begin{figure}[t]
\includegraphics[width=0.6\linewidth]{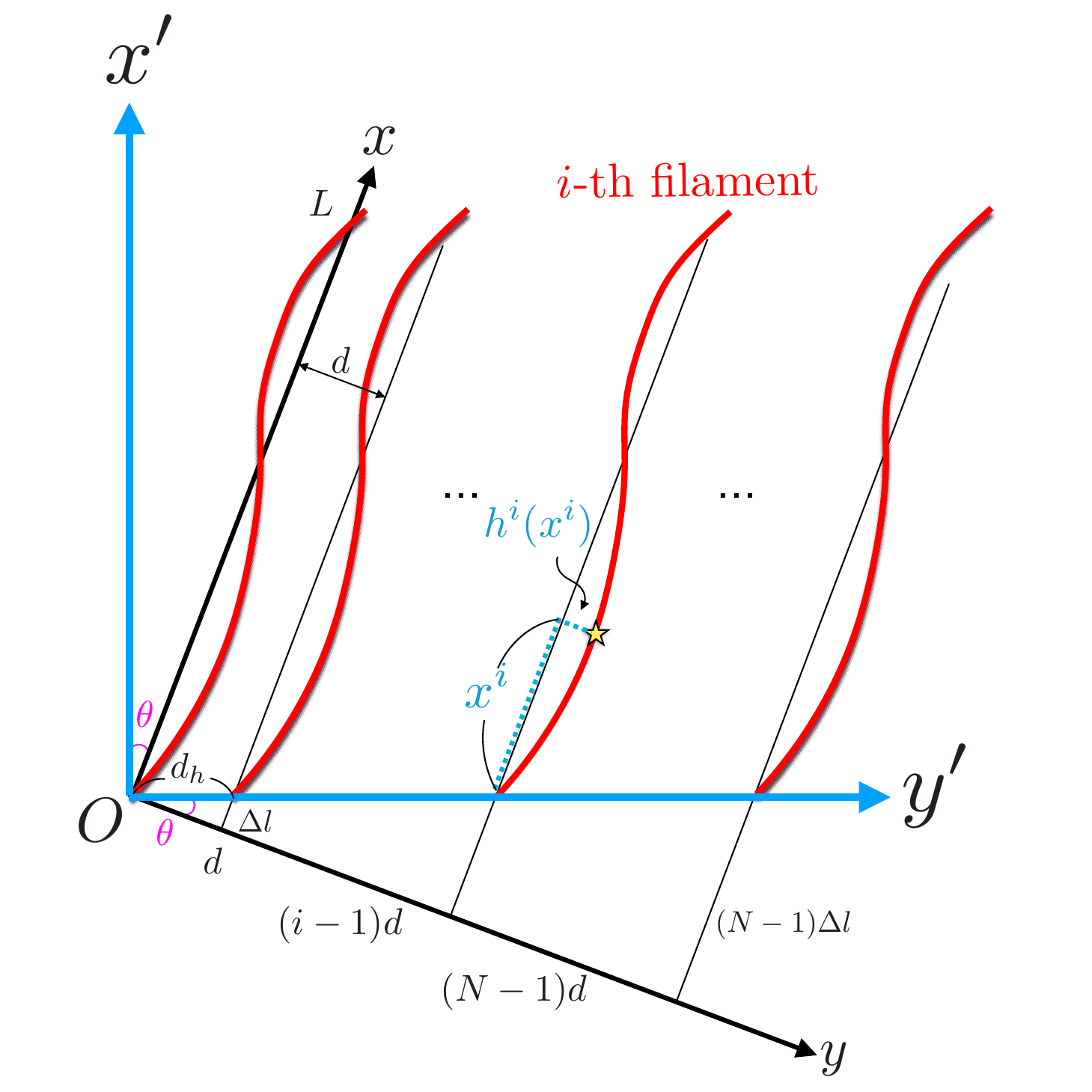}
\caption{Coordinate system used to study the beat dynamics of $N$ filaments aligned in parallel to the $x$-axis with a finite tilt angle $\theta(=\angle x' O x=\angle y'Oy)$ towards the boundary line ($\overline{Oy'}$). 
Here, 
$\Delta l=d\tan{\theta}$ is the size of mismatch. 
The head position of the first filament ($i=1$) is set to the origin $(0,0)$.  
}
\label{fig1}
\end{figure}

{\it Theoretical model.} 
We consider that a 1D array of $N$ slender filaments with radius $a$ and length $L$ are aligned along the $y'$-axis, separated by $d$ ($a\ll d\ll L$), with a tilt angle $\theta$ against $x'$-axis (see Fig.~\ref{fig1}). 
%
The force-balance equation for a filament coupled with its neighbors and the bounding wall is written in a non-dimensionalized form (see Supplemental Materials for details): 
\begin{align}
 \partial_th^i(x^i,t)=
  \mathcal{N}(h^i)+  \mathcal{F}_{\rm HI}(h^{i})+\mathcal{F}_{\rm SI}(h^{i})+\mathcal{F}_{\rm WI}(h^i)
 \label{eqn:main1}
\end{align}
with the boundary conditions, $h^i(0,t)=\partial_x^2h^i(x,t)|_{x^i=0}=0$ and 
$\partial_x^2h^i(x,t)|_{x^i=L}=\partial_x^3h^i(x,t)|_{x^i=L}=0$.   
Here $h^i(x^i,t)$ is the transversal displacement at position $x^i$  along the centerline of $i$th filament and at time $t$. 
In $ \mathcal{N}(h^i)=[-c\partial_x h^i - 2  \partial^2_x h^i+ \left(\partial^2_x h^i\right)^3] - \partial^4_x h^i$,  
the three terms in the parenthesis model the effect of an active force originating from a mechanical feedback of molecular motors~\cite{heuser2009dynein,julicher1997spontaneous}. 
The parameter $c$, denoting a dimensionless speed of wave along the centerline, reflects the filament activity. 

To make the hydrodynamic forces, $\mathcal{F}_{\rm HI}(h^i)$, 
on the $i$th filament interpretable, we consider the resistive force theory~\cite{gray1955propulsion,goldstein2016elastohydrodynamic}, such that 
\begin{align}
\mathcal{F}_{\rm HI}(h^{i})=\sum\limits_{j\neq i}f^{j\rightarrow i}_{\rm HI}/ \zeta_{\perp}=
\epsilon_h \sum\limits_{j\neq i}\partial_t h^j(x^j,t) 
   \label{eqn:hydro1}
\end{align}
where $x^j$ denotes the position along the $j$th filament. 
The coefficient for the strength of HI is given by~\cite{rotne1969variational,yamakawa1970transport} 
\begin{align}
\epsilon_h=\ln{(d/L)}/\ln{(a/L)}+(1/2)(a/d)^2/\ln{(a/L)}. 
\label{eqn:eh}
\end{align}
In our system ($L=100a$, $d=10a$), $\epsilon_h=0.4989$.  
The terms $\mathcal{F}_{\rm SI}(h^{i})$ and $\mathcal{F}_{\rm WI}(h^i)$ 
account for the inter-filament repulsion and the interaction with the bounding wall, respectively. 
SIs in Eq.~\ref{eqn:main1} is necessary in stabilizing 
the in-phase synchronization against the anti-phase synchronization under hydrodynamic coupling, especially for a filament array satisfying the condition $d<\langle \delta h\rangle$ where $\langle\delta h\rangle \equiv \left[(1/T)\int_0^Tdt(1/L)\int_0^Ldx(h(x,t)-\langle h\rangle)^2\right]^{1/2}$ (see SM Text and Fig.~S1). 
We study the collective dynamics of $\{h^i(x^i,t)|i=1,2,\ldots,N\}$ by numerically solving Eq.\ref{eqn:main1}. 

In what follows, we first characterize the single filament dynamics and  next explore how the inter-filament coupling 
contributes to the formation of  MCWs.

\begin{figure}[t!]
    \includegraphics[width=1.0\linewidth]{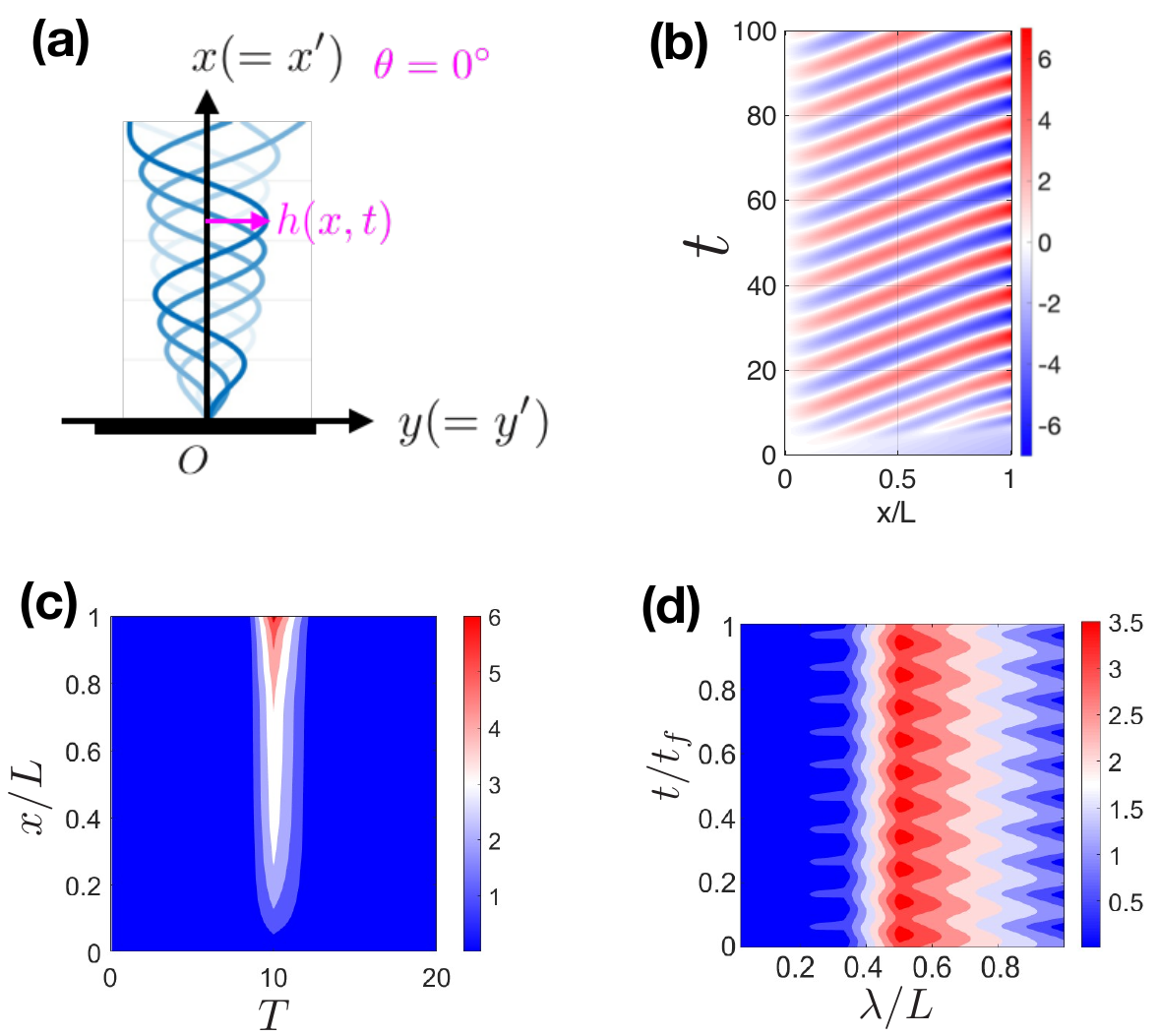}
       \caption{Dynamics of a single filament at $\theta=0^\circ$ (see  SM Movie 1). 
       (a) Filament configurations at $c=1$ in increasing time
       (pale to dark).  
        (b) Kymograph ($h(x,t)$) depicting the spatiotemporal evolution. 
       (c), (d) Power spectra of $h(x,t)$ in time and space:  
		(c) $S(x,T)$, and (d) 
		$F(\lambda,t)$. 
       }
\label{fig:single}
\end{figure}

{\it Single filament dynamics.}
For a single filament ($N=1$) without tilt ($\theta=0^\circ$), the wall interaction (WI) is minimal, which reduces  
Eq.~\ref{eqn:main1} to 
$\partial_th(x,t)= \mathcal{N}(h)$ and yields  
a solution of $h(x,t)\equiv h_o(x,t)$ for $t\gg 1$. 
We use $h_o(x,t)$ as a reference throughout the paper (Fig.~\ref{fig:single}(b), see  SM Movie 1). 
One of the key features of $h_o(x,t)$ is that it preserves LR symmetry
such that both $h_o$ and $-h_o$ are the solution of Eq.\ref{eqn:main1}, 
while displaying a growing amplitude from the head to tail (Fig.~\ref{fig:single}(a)). 
The $x$-dependent beat period, $T(x)$, 
can either be read off from the kymograph (Fig.~\ref{fig:single}(b))  satisfying $h(x,t)\approx h(x,t+T(x))$,  
or be obtained as   
$T(x)=\argmax_TS(x,T)$  
from the power spectrum  
$S(x,T)\sim \int h(x,t)e^{-2\pi i t/T}dt$. 
The spectrum $S(x,T)$ (Fig.~\ref{fig:single}(c)) 
shows a stripe at $T(x)\approx10\equiv T_o$, effectively independent of $x$. 
Similarly, the wavelength at time $t$ is either estimated through $h(x,t)\approx h(x+\lambda(t),t)$ or is 
obtained as $\lambda(t)=\argmax_\lambda F(\lambda,t)$
from the power spectrum 
$F(\lambda,t)\sim \int_0^L h(x,t)e^{2\pi i x/\lambda}dx$ (Fig.~\ref{fig:single}(d)). 
The mean wavelength of a filament for $c=1$ is 
$\lambda_o=0.5L$ with $L=20$. 
We confirm that the wave speed along the filament body calculated from  $\lambda_o=10$ and $T_o=10$ is identical to the input wave speed ($c=1$), i.e., $v_c=\lambda_o/T_o=c=1$.

\begin{figure}[t]
	\includegraphics[width=\linewidth]{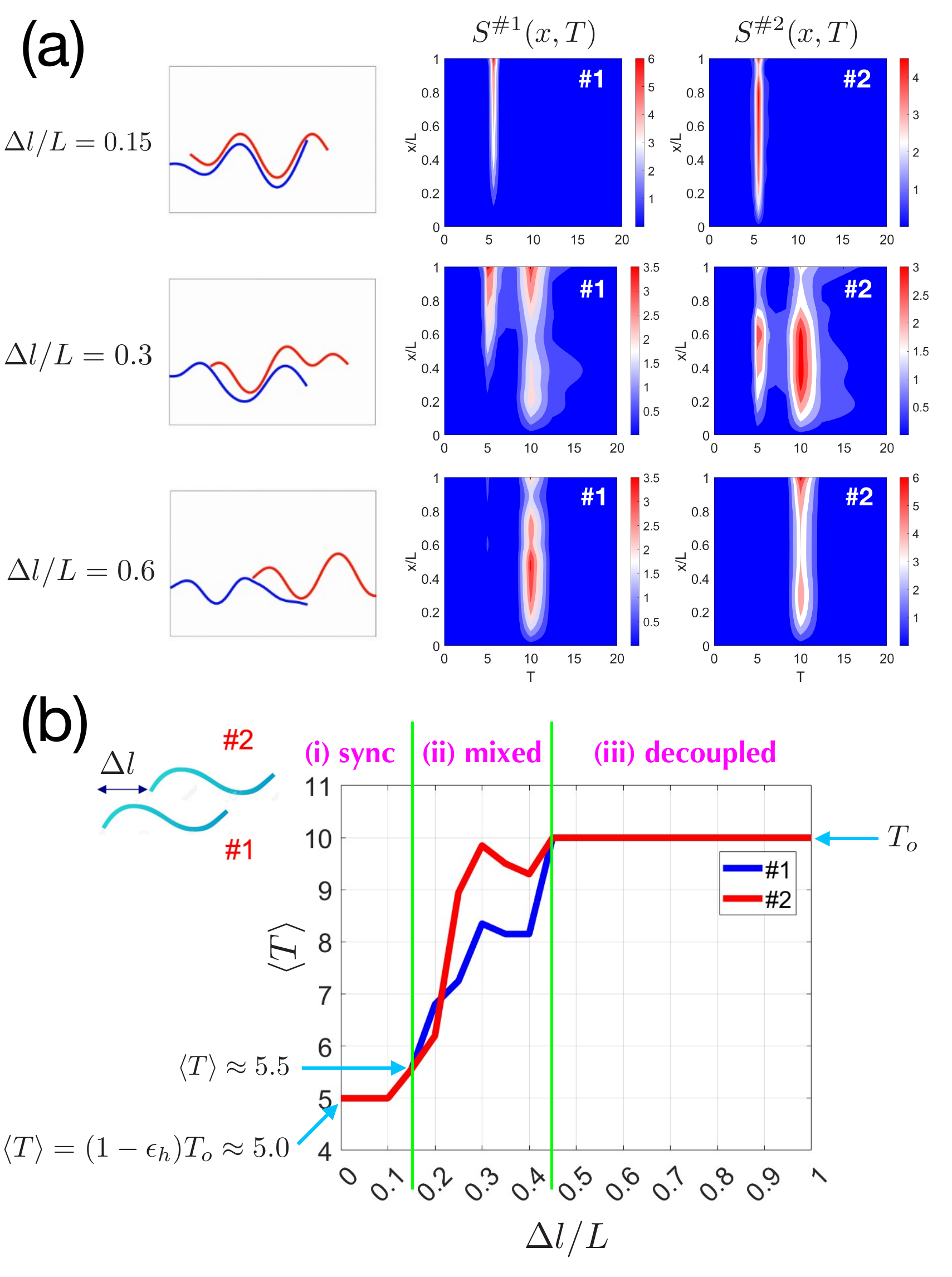}
	\caption{Dynamics of two interacting filaments with increasing mismatch. 
	(a) Configurations of the filaments (\#1 (blue) and \#2 (red)) and their power spectra, $S^{\#1}(x,T)$ and $S^{\#2}(x,T)$, calculated for 
	$\Delta l/L=0.15$, $0.3$, and $0.6$, and $t\gg 1$.  
	(b) Mean undulation period of each filament, $\langle T\rangle=\frac{1}{L}\int_0^LT(x)dx$, as a function of $\Delta l/L$. See SM Movie 2. 
	}
	\label{fig:twomisalign}
\end{figure}

{\it Dynamics of interacting filaments. }
When multiple filaments interact with each other, additional force terms, $\mathcal{F}_{\rm HI}$, $\mathcal{F}_{\rm SI}$, and $\mathcal{F}_{\rm WI}$, come into play. 
The numerical solutions of Eq.\ref{eqn:main1} 
suggest that the coefficient for HIs is renormalized from $\epsilon_h$ to $\epsilon^{\rm eff}$ in the presence of SI and WI. 
The filaments that tilt at an angle 
$\theta$ towards the bounding wall (Fig.~\ref{fig1}) 
do not fully overlap, creating a mismatch, whose size  
$\Delta l$ is related to $\theta$ as $\Delta l=d\tan\theta$ (Fig.~\ref{fig1}). 
With the resistive force theory that enables to replace the hydrodynamic coupling with the velocity of transversal displacement (Eq.~\ref{eqn:hydro1}), the force-balance equation for each filament in the array can be written as  
\begin{align}
\partial_t h^1(x,t)&=\mathcal{N}(h^1)+\epsilon^{\rm eff}\partial_th^{2}(x-\Delta l,t)\nonumber\\
\partial_t h^i(x,t)&=\mathcal{N}(h^i)\nonumber\\
&+\epsilon^{\rm eff}\left[\partial_th^{i-1}(x+\Delta l,t)+\partial_th^{i+1}(x-\Delta l,t)\right]\nonumber\\ 
\partial_t h^{N}(x,t)&=\mathcal{N}(h^N)+\epsilon^{\rm eff}\partial_th^{N-1}(x+\Delta l,t).
\label{eqn:PDE}
\end{align} 
The segment of $(i\mp1)$th filament located at $x\pm\Delta l$ 
interacts with the segment of $i$th filament at $x$.  
When the filaments beat collectively with $\Delta l>0$, 
the transversal displacements of the interacting segments satisfy the inequalities, 
$|h^{i-1}(x+\Delta l,t)|>|h^{i}(x,t)|>|h^{i+1}(x-\Delta l,t)|$. 
Thus, we assume that the transversal displacements of the neighboring filaments satisfy the relation $h^{i\pm1}(x\mp\Delta l,t)\approx b_{\pm}(x,\Delta l)h^{i}(x,t)$, where $b_\pm(x,\Delta l)$
are the factors modulating the transversal displacement at $x$. 

For $\Delta l=0$, the filaments beat in full synchrony, satisfying $h^1\approx  \cdots \approx  h^N\approx h_{\rm sync}$, and do not alter the waveform of adjacent filament ($b_{\pm}(x,0)\approx 1$). 
Summing up the equations ($i=1,\ldots, N$) in Eq.~\ref{eqn:PDE} and dividing it by $N$, one obtains an evolution equation for $h_{\rm sync}(x,t)$,    
$\left[1-2\left(1-1/N\right)\epsilon^{\rm eff}\right]\partial_t h_{\rm sync}(x,t)=\mathcal{N}(h_{\rm sync})$.   
Thus, it can be considered either that $h_{\rm sync}(x,t)$ is mapped to $h_o(x,t)$ by rescaling the time as 
$h_{\rm sync}\left(x,t\right)=h_o\left(x,\frac{t}{1-2(1-1/N)\epsilon^{\rm eff}}\right)$, or that the drag coefficient is reduced from $1$ to 
$\left[1-2\left(1-1/N\right)\epsilon^{\rm eff}\right]$. 
This argument can be generalized to the case with $\Delta l\neq 0$, offering the mean undulation period for an array of $N$ filaments,    
\begin{align}
\langle T_{\theta} \rangle(N) = \left[1-2\bar{b}(\Delta l)\left(1-\frac{1}{N}\right)\epsilon^{\rm eff}\right]T_o   
\label{eqn:T_theta}
\end{align}
with $\bar{b}(\Delta l)\equiv \frac{1}{2L}\int_0^L[b_{-}(x,\Delta l)+b_{+}(x,\Delta l)]dx$. 

For a pair of interacting filaments, we study the detailed effect of misalignment  on their dynamics. 
With increasing $\Delta l$, three different  behaviors emerge: (i) full synchronization; (ii) partial synchronization; (iii) decoupled dynamics. 
Fig.~\ref{fig:twomisalign}(a) displays typical filament configurations and their power spectra $S(x,T)$ for three cases of $\Delta l$.

(i) Sync regime ($0\leq \Delta l/L\leq 0.15$): 
The filament dynamics are fully synchronized in the region where the bodies of two filaments overlap, and there is a minor distinction between $S^{\#1}(x,T)$ and $S^{\#2}(x,T)$. 
The corresponding mean period of undulation increases from $\langle T\rangle=(1-\epsilon_h)T_o\approx 5.0$ to $\approx 5.5$ (see Fig.~\ref{fig:twomisalign}(b)). 
This slight increase is due to the reduced size of overlap between the two filaments.  
In fact, for a filament with a fixed wave speed, 
both $\lambda_o$ and $T_o$ tend to increase with decreasing filament size ($L$)~\cite{goldstein2016elastohydrodynamic}. 

(ii) Mixed regime ($0.15<\Delta l/L<0.45$): 
The tail part of the filament \#1, whose transversal fluctuations is greater than that of head part, interacts with the head part of the filament \#2, engendering a mismatch between $\partial_th^{1}(x^1,t)$ and $\partial_th^{2}(x^2,t)$.  
Since the strength of HIs grows with the amplitude of transversal fluctuation for a fixed oscillatory period, i.e., 
$|f_{\rm HI}^{j \rightarrow i}(x,t)|\propto |h^j(x,t)|$ (Eq.~\ref{eqn:hydro1}),  
the mismatch leads to the LR symmetry breaking of HIs,  $f_{\rm HI}^{2\rightarrow 1}\neq f_{\rm HI}^{1\rightarrow 2}$.  
In this regime, 
the synchronization and decoupling of filament dynamics coexist, and  $S^{\#1}(x,T)$ and $S^{\#2}(x,T)$, exhibiting two main peaks at $T=(1-\epsilon_h)T_o\approx 5$ and $T=T_o\approx 10$, clearly differ from each other, suggestive of the broken LR symmetry in the beat dynamics of the two. 

\begin{figure*}[th!]
	\includegraphics[width=\linewidth]{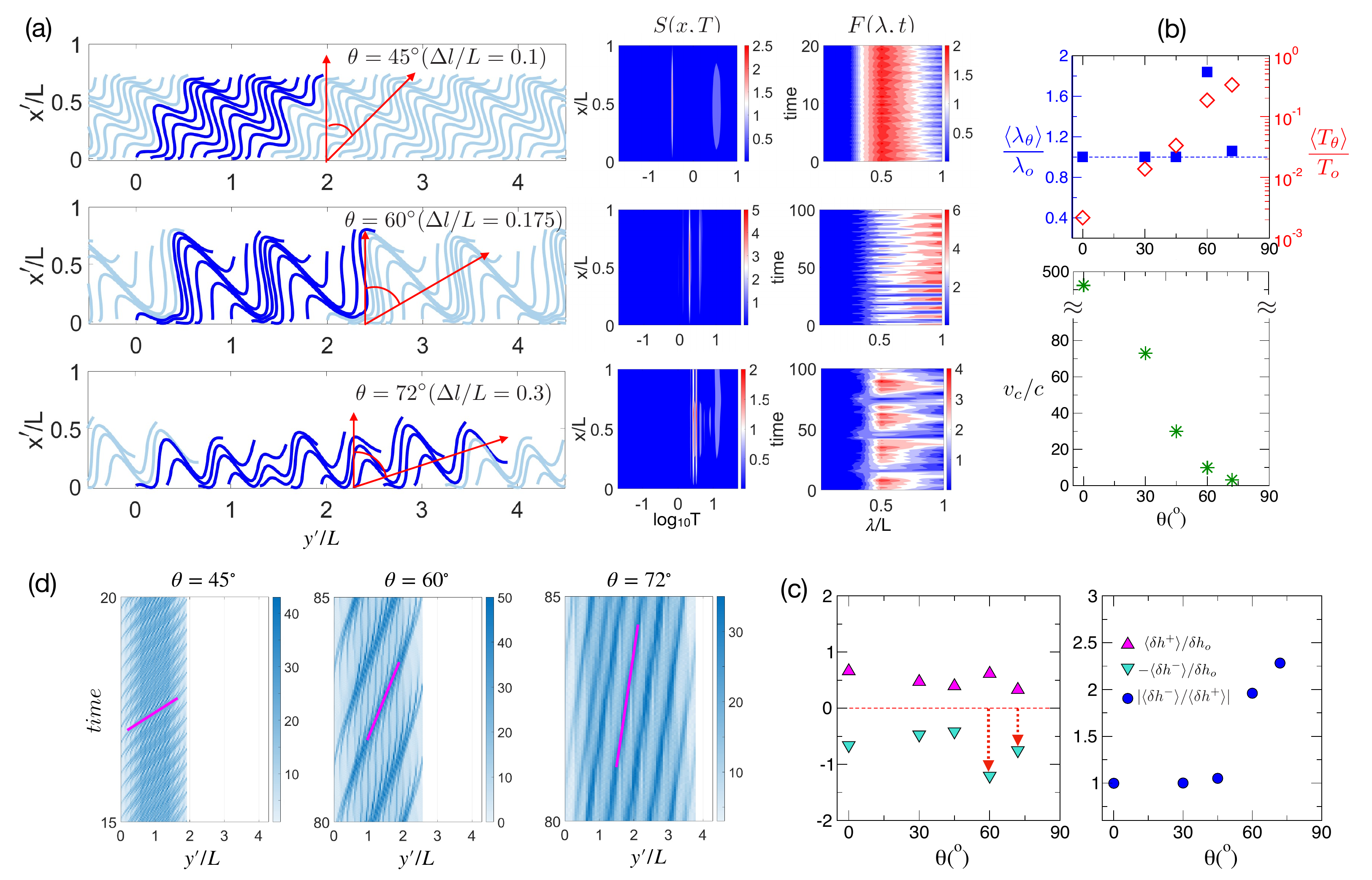}
	\caption{
MCWs generated under periodic boundary (SM Movie 3). 
	(a) MCWs generated at three $\theta$ values and the power spectra, $S(x,T)$ and $F(\lambda,t)$.  
	(b) $\langle\lambda\rangle$, $\langle T\rangle$, and $v_c$ for the individual filaments in the array with increasing $\theta$. 
	(c) The mean transversal fluctuations of filaments with increasing $\theta$: $\langle\delta h^+\rangle$, and $\langle\delta h^-\rangle$, where $\langle \delta h^\pm\rangle$ is the root mean squared positive (or negative) fluctuations.  
	(left) The large amplitudes of $\langle \delta h^-\rangle$ at $\theta=60^\circ$ and $72^\circ$ is highlighted by the red arrows.	
	(right) The ratio of left/right amplitudes, $|\langle \delta h^-\rangle/\langle \delta h^+\rangle|$. 
	(d) Spatiotemporal evolution of mass density projected on the $y'$-axis. 
	 The slopes of stripes, marked with magenta lines, are used to estimate the speed of MCWs. 
       	}
	\label{fig:N10pBC}
\end{figure*}

(iii) Decoupled regime ($0.45<\Delta l/L\leq 1$): The peak of $S(x,T)$ is unique  
at $T\approx T_o$. 
Although 40 $\%$ of the filament body is subject to HIs for $\Delta l/L=0.6$, the dynamics of two filaments are effectively decoupled, 
and the waveform is restored to $h_o(x,t)$. 
An increasing mismatch ($\Delta l$) weakens the strength of HIs (or reduces $\bar{b}(\Delta l)$), and it consequently tends to increase $\langle T\rangle$ from the case with zero mismatch 
to the case with full decoupling, i.e., $(1-\epsilon_h)T_o\rightarrow T_o$. 
From the regime (i) to (iii), 
the relative weight of the two main peaks in $S(x,T)$ 
shifts from $T\approx (1-\epsilon_h)T_o$ to $T\approx T_o$ and undergoes a first order type discrete transition (Fig.~\ref{fig:twomisalign}(a)).

\begin{figure}[t]
	\includegraphics[width=\linewidth]{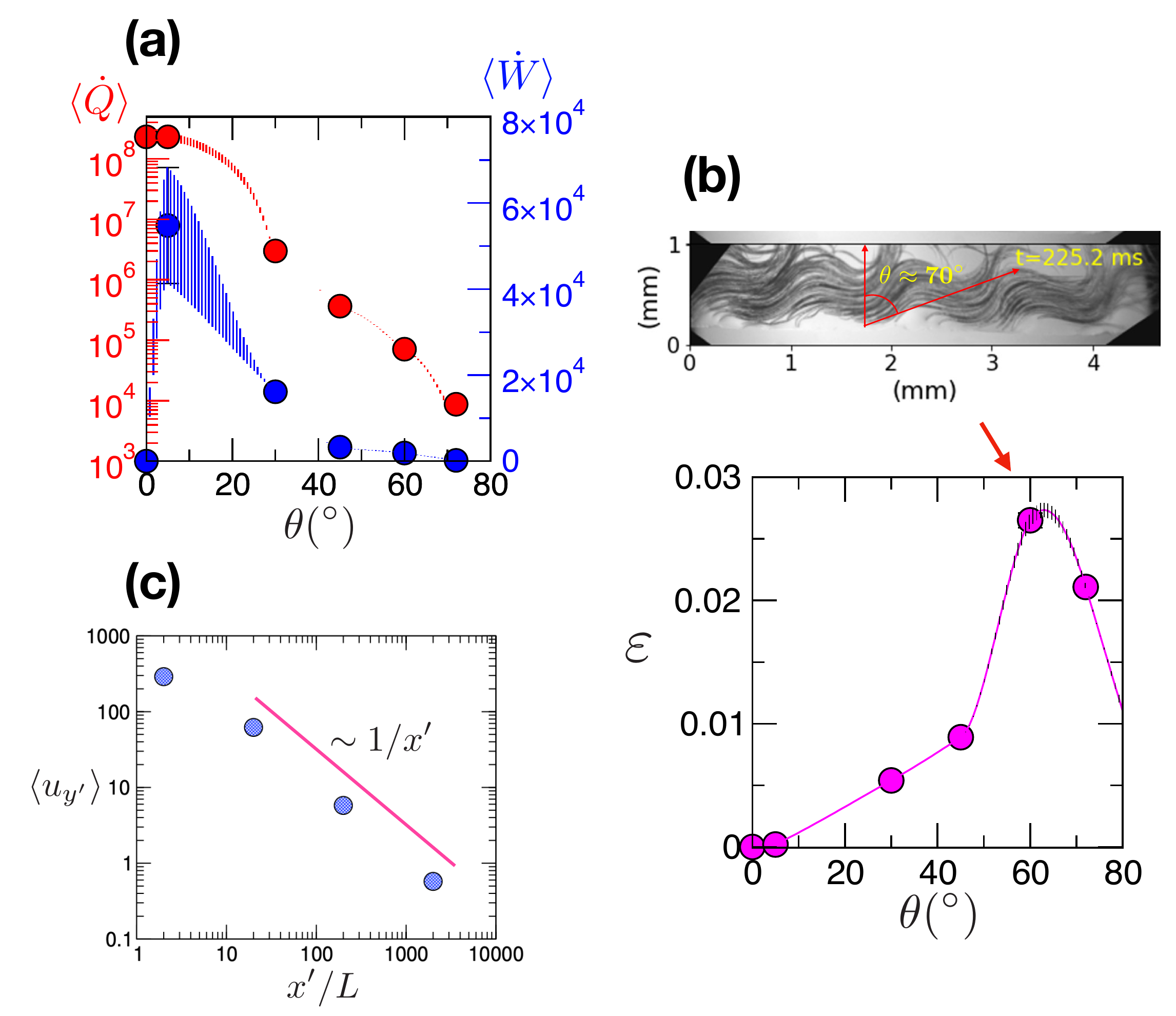}
	\caption{
	(a) Work production and dissipation, and (b) thermodynamic efficiency with varying $\theta$. 
	The inset depicts a nematode array with $\theta\approx 70^\circ$ from the Peshkov \emph{et al.}'s experiment~\cite{peshkov2022synchronized}. 
	(c) The mean velocity of fluid flow at $x'$ calculated for an array of filaments ($N=400$) at $\theta=30^\circ$. 
	}
	\label{fig:fluid}
\end{figure}


We now analyze behaviors of 1D arrays of many filaments ($N=10$) at a fixed separation ($d=10a$) by imposing a periodic boundary condition to eliminate the boundary effect (see SM Text and Fig.~S2 for the case without periodic boundary). 


First, for $\theta=0^\circ$,  
the filaments beat in perfect synchrony at steady states, yielding the 
waveform, 
$h_{\rm sync}(x,t)$, whose wavelength and transversal fluctuation are identical to those of a single filament in isolation ($\langle \lambda \rangle =\lambda_o=0.5L$ and $\langle \delta h\rangle =\delta h_o=0.17L$), yet the time is rescaled as 
$t\rightarrow t/(1-2\epsilon_h)$. 
The oscillatory period is reduced to $\langle T \rangle =2.16\times 10^{-3}T_o$, which substantially accelerates the wave speed to $v_c= 462c$. 
This is in excellent agreement with a theoretical estimate, $\langle T_{0^\circ}\rangle=(1-2\epsilon_h)T_o\approx 2.20\times 10^{-3}T_o$ and $v_c= c/(1-2\epsilon_h)\approx 460c$, where we have used  Eq.~\ref{eqn:T_theta} with $\bar{b}(\Delta l=0)=1$, $\epsilon_h\approx 0.4989$ (Eq.~\ref{eqn:eh}) and $N\rightarrow \infty$. 
The filaments beat without phase or time lag, retaining the LR symmetry. 

Next, for $\theta>0^\circ$, the waveforms for all the filaments are modified from $h_{0^\circ}(x,t)$ to $h_{\theta}(x,t)$, while the individual filaments collectively undulate with a constant time lag $\tau_{\theta}$, generating MCWs that travel across the array of filaments ($y'$-axis), which differ from the body waves traveling along the centerline ($x$-axis) of each filament at speed $v_c$. 
%
$\langle T_{\theta}\rangle$, calculated by analyzing $S(x,T)$, 
displays a clear increase with $\theta$ ($S(x,T)$ in Figs.~\ref{fig:N10pBC}(a) and ~\ref{fig:N10pBC}(b)). 
Although $\langle\lambda_\theta\rangle$ 
changes non-monotonically, maximized at $\theta=60^\circ$, 
$v_c=\langle\lambda_\theta\rangle/\langle T_\theta\rangle$ decreases with $\theta$ (Fig.~\ref{fig:N10pBC}(b)). 

In light of the two-filament dynamics (Fig.~\ref{fig:twomisalign}(b)), 
the cases with $\theta=60^{\circ}$ 
and $72^{\circ}$ (Fig.~\ref{fig:N10pBC}(a)) belong to the mixed regime, and 
the deviation of the wave characteristics from those of a single filament is evident (Fig.~\ref{fig:N10pBC}(b)).   
Importantly, for $\theta=60^{\circ}$, the transversal displacement of filaments is significantly skewed towards the left side of the  centerline, i.e., $|\langle\delta h^-\rangle|>|\langle\delta h^+\rangle|$ (see the red arrows in Fig.~\ref{fig:N10pBC}(c)). 
Accordingly, the hydrodynamic coupling becomes stronger on the left side  than on the right, breaking the LR symmetry of HIs. 

The collective waveform (MCW)
propagates along the array with a constant time lag $\tau_\theta$  between adjacent filaments. 
The $\tau_\theta$ is 
the time required for a position of a filament to travel from $x-\Delta l$ to $x$ and mimic the movement of the adjacent filament, whose head positions are separated by $d_h=\Delta l/\sin\theta$. Note that $\tau_{\theta}$ is related to $v_c(\theta)$ as $\tau_{\theta}=\Delta l/v_c(\theta)$. 
For $\theta > 0^\circ$, the speed of MCW, $v_M$, relates to $v_c$ as follows. 
\begin{align}
	v_{\rm M}(\theta)=\frac{d_h}{\tau_\theta}=\frac{d_h}{\Delta l}v_c(\theta)=\frac{v_c(\theta)}{\sin{\theta}}. 
	\label{eq:vm}
\end{align} 
For different $\theta$, the speeds of MCWs are:   
$v_{\rm M}^{45^\circ}=42.4$, 
$v_{\rm M}^{60^\circ}=11.5$, 
and 
$v_{\rm M}^{72^\circ}=3.4$. 
Alternatively, the $v_{\rm M}$ can be determined by analyzing the spatiotemporal evolution of mass density projected on the $y'$ axis, 
$\rho(y',t)$. 
As clearly shown in Fig.~\ref{fig:N10pBC}(d), $\rho(y',t)$ is periodic in space and time, $\rho(y',t)\approx \rho(y'+\Delta y'^\ast,t)$ and $\rho(y',t)\approx \rho(y',t+\Delta t^\ast)$, which enables to estimate the $v_{\rm M}$ by using $v_{\rm M}^\ast=\Delta y'^\ast/\Delta t^\ast$. 
The estimated speeds, 
$v_{\rm M}^{\ast,45^{\circ}}\approx 40$, $v_{\rm M}^{\ast,60^{\circ}}\approx 10$ and $v_{\rm M}^{\ast,72^{\circ}}\approx 4$, 
agree well with the values of $v_{\rm M}$ obtained using Eq.~\ref{eq:vm}.

{\it Viscous dissipation, work production, and thermodynamic efficiency. }
The mean viscous dissipation per filament ($\langle\dot{Q}\rangle$)
and the work production rate per filament ($\langle\dot{W}\rangle$), which 
enhances the fluid flow along the $y'$-direction,  
can be evaluated from the waveforms of filaments $\{h^i(x^i,t)\}$ (see Eqs. S23 and S24). 
While $\langle\dot{W}\rangle$ is maximized at $\theta\simeq 5^\circ$ (Fig.~\ref{fig:fluid}(a)), 
the extent of the mechanical work 
relative to the total dissipation, quantified by 
$\varepsilon(\theta) =\langle\dot{W}\rangle/\langle\dot{Q}\rangle$, is maximized in the range of $\theta\approx 60^\circ-70^\circ$ (Fig.~\ref{fig:fluid}(b)). 
Remarkably, 
the array of \emph{T. aceti}~\cite{peshkov2022synchronized} adopt effectively the same range of tilt angle ($\theta\approx 70^\circ$, see the inset of Fig.~\ref{fig:fluid}(b)), suggesting that their configuration is spontaneously adjusted to maximize the thermodynamic efficiency ($\varepsilon$). 

Lastly, the fluid flow generated by the MCW is calculated by using the linear response between the propulsion force and the fluid velocity associated through a hydrodynamic tensor (Eq.~S5, see SM Text for  details). 
Fig.~\ref{fig:fluid}(c) shows that 
the velocity of fluid flow along $y'$-axis 
decreases with the distance from the $y'$-axis as $u_{y'}\sim 1/x'$ for $x'\gg 1$, 
reflecting a $1/r$-decay characteristic of the fundamental solution of the hydrodynamic tensor (Eq.~S5). 
 
{\it Concluding Remarks. }
The modulation of filament waveform is one of the key features of MCWs observed in the \emph{T. aceti} experiment~\cite{peshkov2022synchronized}. 
To explore MCWs, previous studies  
adopted a phase reduction approach, representing a beat dynamics by means of a frequency-locked phase oscillator
~\cite{nakao2016phase,kawamura2018phase,man2020multisynchrony,quillen2021metachronal,quillen2023robust}. 
Such a strategy may prove effective for perfectly aligned filaments whose waveform remains unaltered in synchrony (see Fig.~S2)~\cite{guo2018bistability,young2009hydrodynamic,vilfan2006hydrodynamic}; however, it overlooks 
the physical reality that the wave characteristics of each filament is subject to change upon interacting with its neighbors. 
Our study solving the elastohydrodynamic equation 
clarifies that the full evolution of waveforms is essential to evaluate the energetics of a system as well as to decipher the physics giving rise to MCWs.  

Unlike the asymmetric beat dynamics of cilium, the dynamics of a nematode is symmetric. 
Generation of MCWs from the nematode population appears unlikely at first sight. 
Disruption of the filament synchronization caused by the tilt-induced misalignment and a constant time lag are the essential mechanism for the formation of MCWs.
Importantly, the energetically most efficient MCWs for the fluid circulation are generated at large tilts ($\theta=60^\circ-70^\circ$).

\begin{acknowledgments}
This study was supported by the KIAS individual Grants, CG085701 (NJ), CG076002 (WKK), and CG035003 (CH). 
We thank the Center for Advanced Computation in KIAS for providing the computing resources.
\end{acknowledgments}


\begin{thebibliography}{47}%
\makeatletter
\providecommand \@ifxundefined [1]{%
 \@ifx{#1\undefined}
}%
\providecommand \@ifnum [1]{%
 \ifnum #1\expandafter \@firstoftwo
 \else \expandafter \@secondoftwo
 \fi
}%
\providecommand \@ifx [1]{%
 \ifx #1\expandafter \@firstoftwo
 \else \expandafter \@secondoftwo
 \fi
}%
\providecommand \natexlab [1]{#1}%
\providecommand \enquote  [1]{``#1''}%
\providecommand \bibnamefont  [1]{#1}%
\providecommand \bibfnamefont [1]{#1}%
\providecommand \citenamefont [1]{#1}%
\providecommand \href@noop [0]{\@secondoftwo}%
\providecommand \href [0]{\begingroup \@sanitize@url \@href}%
\providecommand \@href[1]{\@@startlink{#1}\@@href}%
\providecommand \@@href[1]{\endgroup#1\@@endlink}%
\providecommand \@sanitize@url [0]{\catcode `\\12\catcode `\$12\catcode
  `\&12\catcode `\#12\catcode `\^12\catcode `\_12\catcode `\%12\relax}%
\providecommand \@@startlink[1]{}%
\providecommand \@@endlink[0]{}%
\providecommand \url  [0]{\begingroup\@sanitize@url \@url }%
\providecommand \@url [1]{\endgroup\@href {#1}{\urlprefix }}%
\providecommand \urlprefix  [0]{URL }%
\providecommand \Eprint [0]{\href }%
\providecommand \doibase [0]{https://doi.org/}%
\providecommand \selectlanguage [0]{\@gobble}%
\providecommand \bibinfo  [0]{\@secondoftwo}%
\providecommand \bibfield  [0]{\@secondoftwo}%
\providecommand \translation [1]{[#1]}%
\providecommand \BibitemOpen [0]{}%
\providecommand \bibitemStop [0]{}%
\providecommand \bibitemNoStop [0]{.\EOS\space}%
\providecommand \EOS [0]{\spacefactor3000\relax}%
\providecommand \BibitemShut  [1]{\csname bibitem#1\endcsname}%
\let\auto@bib@innerbib\@empty
\bibitem [{\citenamefont {Couzin}(2007)}]{couzin2007collective}%
  \BibitemOpen
  \bibfield  {author} {\bibinfo {author} {\bibfnamefont {I.}~\bibnamefont
  {Couzin}},\ }\bibfield  {title} {\bibinfo {title} {Collective minds},\
  }\href@noop {} {\bibfield  {journal} {\bibinfo  {journal} {Nature}\ }\textbf
  {\bibinfo {volume} {445}},\ \bibinfo {pages} {715} (\bibinfo {year}
  {2007})}\BibitemShut {NoStop}%
\bibitem [{\citenamefont {Elgeti}\ \emph {et~al.}(2015)\citenamefont {Elgeti},
  \citenamefont {Winkler},\ and\ \citenamefont {Gompper}}]{elgeti2015physics}%
  \BibitemOpen
  \bibfield  {author} {\bibinfo {author} {\bibfnamefont {J.}~\bibnamefont
  {Elgeti}}, \bibinfo {author} {\bibfnamefont {R.~G.}\ \bibnamefont
  {Winkler}},\ and\ \bibinfo {author} {\bibfnamefont {G.}~\bibnamefont
  {Gompper}},\ }\bibfield  {title} {\bibinfo {title} {Physics of
  microswimmers—single particle motion and collective behavior: a review},\
  }\href@noop {} {\bibfield  {journal} {\bibinfo  {journal} {Rep. Prog. Phys.}\
  }\textbf {\bibinfo {volume} {78}},\ \bibinfo {pages} {056601} (\bibinfo
  {year} {2015})}\BibitemShut {NoStop}%
\bibitem [{\citenamefont {Cavagna}\ and\ \citenamefont
  {Giardina}(2014)}]{cavagna2014bird}%
  \BibitemOpen
  \bibfield  {author} {\bibinfo {author} {\bibfnamefont {A.}~\bibnamefont
  {Cavagna}}\ and\ \bibinfo {author} {\bibfnamefont {I.}~\bibnamefont
  {Giardina}},\ }\bibfield  {title} {\bibinfo {title} {Bird flocks as condensed
  matter},\ }\href@noop {} {\bibfield  {journal} {\bibinfo  {journal} {Annu.
  Rev. Condens. Matter Phys.}\ }\textbf {\bibinfo {volume} {5}},\ \bibinfo
  {pages} {183} (\bibinfo {year} {2014})}\BibitemShut {NoStop}%
\bibitem [{\citenamefont {Moussa{\"\i}d}\ \emph {et~al.}(2011)\citenamefont
  {Moussa{\"\i}d}, \citenamefont {Helbing},\ and\ \citenamefont
  {Theraulaz}}]{moussaid2011simple}%
  \BibitemOpen
  \bibfield  {author} {\bibinfo {author} {\bibfnamefont {M.}~\bibnamefont
  {Moussa{\"\i}d}}, \bibinfo {author} {\bibfnamefont {D.}~\bibnamefont
  {Helbing}},\ and\ \bibinfo {author} {\bibfnamefont {G.}~\bibnamefont
  {Theraulaz}},\ }\bibfield  {title} {\bibinfo {title} {How simple rules
  determine pedestrian behavior and crowd disasters},\ }\href@noop {}
  {\bibfield  {journal} {\bibinfo  {journal} {Proc. Natl. Acad. Sci. U. S. A.}\
  }\textbf {\bibinfo {volume} {108}},\ \bibinfo {pages} {6884} (\bibinfo {year}
  {2011})}\BibitemShut {NoStop}%
\bibitem [{\citenamefont {Lee}\ \emph {et~al.}(2018)\citenamefont {Lee},
  \citenamefont {Hyeon},\ and\ \citenamefont {Jo}}]{lee2018PRE}%
  \BibitemOpen
  \bibfield  {author} {\bibinfo {author} {\bibfnamefont {S.}~\bibnamefont
  {Lee}}, \bibinfo {author} {\bibfnamefont {C.}~\bibnamefont {Hyeon}},\ and\
  \bibinfo {author} {\bibfnamefont {J.}~\bibnamefont {Jo}},\ }\bibfield
  {title} {\bibinfo {title} {Thermodynamic uncertainty relation of interacting
  oscillators in synchrony},\ }\href@noop {} {\bibfield  {journal} {\bibinfo
  {journal} {Phys. Rev. E}\ }\textbf {\bibinfo {volume} {98}},\ \bibinfo
  {pages} {032119} (\bibinfo {year} {2018})}\BibitemShut {NoStop}%
\bibitem [{\citenamefont {Hong}\ \emph {et~al.}(2020)\citenamefont {Hong},
  \citenamefont {Jo}, \citenamefont {Hyeon},\ and\ \citenamefont
  {Park}}]{hong2020JSM}%
  \BibitemOpen
  \bibfield  {author} {\bibinfo {author} {\bibfnamefont {H.}~\bibnamefont
  {Hong}}, \bibinfo {author} {\bibfnamefont {J.}~\bibnamefont {Jo}}, \bibinfo
  {author} {\bibfnamefont {C.}~\bibnamefont {Hyeon}},\ and\ \bibinfo {author}
  {\bibfnamefont {H.}~\bibnamefont {Park}},\ }\bibfield  {title} {\bibinfo
  {title} {Thermodynamic cost of synchronizing a population of beating cilia},\
  }\href@noop {} {\bibfield  {journal} {\bibinfo  {journal} {J. Stat. Mech.}\
  ,\ \bibinfo {pages} {074001}} (\bibinfo {year} {2020})}\BibitemShut {NoStop}%
\bibitem [{\citenamefont {Taylor}(1951)}]{taylor1951analysis}%
  \BibitemOpen
  \bibfield  {author} {\bibinfo {author} {\bibfnamefont {G.~I.}\ \bibnamefont
  {Taylor}},\ }\bibfield  {title} {\bibinfo {title} {Analysis of the swimming
  of microscopic organisms},\ }\href@noop {} {\bibfield  {journal} {\bibinfo
  {journal} {Proc. R. Soc. A: Math. Phys. Eng. Sci.}\ }\textbf {\bibinfo
  {volume} {209}},\ \bibinfo {pages} {447} (\bibinfo {year}
  {1951})}\BibitemShut {NoStop}%
\bibitem [{\citenamefont {Taylor}(1952)}]{taylor1952analysis}%
  \BibitemOpen
  \bibfield  {author} {\bibinfo {author} {\bibfnamefont {G.~I.}\ \bibnamefont
  {Taylor}},\ }\bibfield  {title} {\bibinfo {title} {Analysis of the swimming
  of long and narrow animals},\ }\href@noop {} {\bibfield  {journal} {\bibinfo
  {journal} {Proc. R. Soc. A: Math. Phys. Eng. Sci.}\ }\textbf {\bibinfo
  {volume} {214}},\ \bibinfo {pages} {158} (\bibinfo {year}
  {1952})}\BibitemShut {NoStop}%
\bibitem [{\citenamefont {Gueron}\ \emph {et~al.}(1997)\citenamefont {Gueron},
  \citenamefont {Levit-Gurevich}, \citenamefont {Liron},\ and\ \citenamefont
  {Blum}}]{gueron1997cilia}%
  \BibitemOpen
  \bibfield  {author} {\bibinfo {author} {\bibfnamefont {S.}~\bibnamefont
  {Gueron}}, \bibinfo {author} {\bibfnamefont {K.}~\bibnamefont
  {Levit-Gurevich}}, \bibinfo {author} {\bibfnamefont {N.}~\bibnamefont
  {Liron}},\ and\ \bibinfo {author} {\bibfnamefont {J.~J.}\ \bibnamefont
  {Blum}},\ }\bibfield  {title} {\bibinfo {title} {Cilia internal mechanism and
  metachronal coordination as the result of hydrodynamical coupling},\
  }\href@noop {} {\bibfield  {journal} {\bibinfo  {journal} {Proc. Natl. Acad.
  Sci.}\ }\textbf {\bibinfo {volume} {94}},\ \bibinfo {pages} {6001} (\bibinfo
  {year} {1997})}\BibitemShut {NoStop}%
\bibitem [{\citenamefont {Gueron}\ and\ \citenamefont
  {Levit-Gurevich}(1999)}]{gueron1999energetic}%
  \BibitemOpen
  \bibfield  {author} {\bibinfo {author} {\bibfnamefont {S.}~\bibnamefont
  {Gueron}}\ and\ \bibinfo {author} {\bibfnamefont {K.}~\bibnamefont
  {Levit-Gurevich}},\ }\bibfield  {title} {\bibinfo {title} {Energetic
  considerations of ciliary beating and the advantage of metachronal
  coordination},\ }\href@noop {} {\bibfield  {journal} {\bibinfo  {journal}
  {Proc. Natl. Acad. Sci. U. S. A.}\ }\textbf {\bibinfo {volume} {96}},\
  \bibinfo {pages} {12240} (\bibinfo {year} {1999})}\BibitemShut {NoStop}%
\bibitem [{\citenamefont {Niedermayer}\ \emph {et~al.}(2008)\citenamefont
  {Niedermayer}, \citenamefont {Eckhardt},\ and\ \citenamefont
  {Lenz}}]{niedermayer2008synchronization}%
  \BibitemOpen
  \bibfield  {author} {\bibinfo {author} {\bibfnamefont {T.}~\bibnamefont
  {Niedermayer}}, \bibinfo {author} {\bibfnamefont {B.}~\bibnamefont
  {Eckhardt}},\ and\ \bibinfo {author} {\bibfnamefont {P.}~\bibnamefont
  {Lenz}},\ }\bibfield  {title} {\bibinfo {title} {Synchronization, phase
  locking, and metachronal wave formation in ciliary chains},\ }\href@noop {}
  {\bibfield  {journal} {\bibinfo  {journal} {Chaos}\ }\textbf {\bibinfo
  {volume} {18}} (\bibinfo {year} {2008})}\BibitemShut {NoStop}%
\bibitem [{\citenamefont {Gu}\ \emph {et~al.}(2020)\citenamefont {Gu},
  \citenamefont {Boehler}, \citenamefont {Cui}, \citenamefont {Secchi},
  \citenamefont {Savorana}, \citenamefont {De~Marco}, \citenamefont
  {Gervasoni}, \citenamefont {Peyron}, \citenamefont {Huang}, \citenamefont
  {Pane} \emph {et~al.}}]{gu2020magnetic}%
  \BibitemOpen
  \bibfield  {author} {\bibinfo {author} {\bibfnamefont {H.}~\bibnamefont
  {Gu}}, \bibinfo {author} {\bibfnamefont {Q.}~\bibnamefont {Boehler}},
  \bibinfo {author} {\bibfnamefont {H.}~\bibnamefont {Cui}}, \bibinfo {author}
  {\bibfnamefont {E.}~\bibnamefont {Secchi}}, \bibinfo {author} {\bibfnamefont
  {G.}~\bibnamefont {Savorana}}, \bibinfo {author} {\bibfnamefont
  {C.}~\bibnamefont {De~Marco}}, \bibinfo {author} {\bibfnamefont
  {S.}~\bibnamefont {Gervasoni}}, \bibinfo {author} {\bibfnamefont
  {Q.}~\bibnamefont {Peyron}}, \bibinfo {author} {\bibfnamefont {T.-Y.}\
  \bibnamefont {Huang}}, \bibinfo {author} {\bibfnamefont {S.}~\bibnamefont
  {Pane}}, \emph {et~al.},\ }\bibfield  {title} {\bibinfo {title} {Magnetic
  cilia carpets with programmable metachronal waves},\ }\href@noop {}
  {\bibfield  {journal} {\bibinfo  {journal} {Nature Commun.}\ }\textbf
  {\bibinfo {volume} {11}},\ \bibinfo {pages} {2637} (\bibinfo {year}
  {2020})}\BibitemShut {NoStop}%
\bibitem [{\citenamefont {Wang}\ \emph {et~al.}(2024)\citenamefont {Wang},
  \citenamefont {ul~Islam}, \citenamefont {Steur}, \citenamefont {Homan},
  \citenamefont {Aggarwal}, \citenamefont {Onck}, \citenamefont {den Toonder},\
  and\ \citenamefont {Wang}}]{wang2024programmable}%
  \BibitemOpen
  \bibfield  {author} {\bibinfo {author} {\bibfnamefont {T.}~\bibnamefont
  {Wang}}, \bibinfo {author} {\bibfnamefont {T.}~\bibnamefont {ul~Islam}},
  \bibinfo {author} {\bibfnamefont {E.}~\bibnamefont {Steur}}, \bibinfo
  {author} {\bibfnamefont {T.}~\bibnamefont {Homan}}, \bibinfo {author}
  {\bibfnamefont {I.}~\bibnamefont {Aggarwal}}, \bibinfo {author}
  {\bibfnamefont {P.~R.}\ \bibnamefont {Onck}}, \bibinfo {author}
  {\bibfnamefont {J.~M.}\ \bibnamefont {den Toonder}},\ and\ \bibinfo {author}
  {\bibfnamefont {Y.}~\bibnamefont {Wang}},\ }\bibfield  {title} {\bibinfo
  {title} {Programmable metachronal motion of closely packed magnetic
  artificial cilia},\ }\href@noop {} {\bibfield  {journal} {\bibinfo  {journal}
  {Lab on a Chip}\ }\textbf {\bibinfo {volume} {24}},\ \bibinfo {pages} {1573}
  (\bibinfo {year} {2024})}\BibitemShut {NoStop}%
\bibitem [{\citenamefont {Hancock}(1953)}]{hancock1953self}%
  \BibitemOpen
  \bibfield  {author} {\bibinfo {author} {\bibfnamefont {G.}~\bibnamefont
  {Hancock}},\ }\bibfield  {title} {\bibinfo {title} {The self-propulsion of
  microscopic organisms through liquids},\ }\href@noop {} {\bibfield  {journal}
  {\bibinfo  {journal} {Proc. R. Soc. A: Math. Phys. Eng. Sci.}\ }\textbf
  {\bibinfo {volume} {217}},\ \bibinfo {pages} {96} (\bibinfo {year}
  {1953})}\BibitemShut {NoStop}%
\bibitem [{\citenamefont {Yang}\ \emph {et~al.}(2008)\citenamefont {Yang},
  \citenamefont {Elgeti},\ and\ \citenamefont {Gompper}}]{yang2008cooperation}%
  \BibitemOpen
  \bibfield  {author} {\bibinfo {author} {\bibfnamefont {Y.}~\bibnamefont
  {Yang}}, \bibinfo {author} {\bibfnamefont {J.}~\bibnamefont {Elgeti}},\ and\
  \bibinfo {author} {\bibfnamefont {G.}~\bibnamefont {Gompper}},\ }\bibfield
  {title} {\bibinfo {title} {Cooperation of sperm in two dimensions:
  synchronization, attraction, and aggregation through hydrodynamic
  interactions},\ }\href@noop {} {\bibfield  {journal} {\bibinfo  {journal}
  {Phys. Rev. E.}\ }\textbf {\bibinfo {volume} {78}},\ \bibinfo {pages}
  {061903} (\bibinfo {year} {2008})}\BibitemShut {NoStop}%
\bibitem [{\citenamefont {Golestanian}\ \emph {et~al.}(2011)\citenamefont
  {Golestanian}, \citenamefont {Yeomans},\ and\ \citenamefont
  {Uchida}}]{golestanian2011hydrodynamic}%
  \BibitemOpen
  \bibfield  {author} {\bibinfo {author} {\bibfnamefont {R.}~\bibnamefont
  {Golestanian}}, \bibinfo {author} {\bibfnamefont {J.~M.}\ \bibnamefont
  {Yeomans}},\ and\ \bibinfo {author} {\bibfnamefont {N.}~\bibnamefont
  {Uchida}},\ }\bibfield  {title} {\bibinfo {title} {{Hydrodynamic
  synchronization at low Reynolds number}},\ }\href@noop {} {\bibfield
  {journal} {\bibinfo  {journal} {Soft Matter}\ }\textbf {\bibinfo {volume}
  {7}},\ \bibinfo {pages} {3074} (\bibinfo {year} {2011})}\BibitemShut
  {NoStop}%
\bibitem [{\citenamefont {Kim}\ and\ \citenamefont
  {Netz}(2006)}]{kim2006pumping}%
  \BibitemOpen
  \bibfield  {author} {\bibinfo {author} {\bibfnamefont {Y.~W.}\ \bibnamefont
  {Kim}}\ and\ \bibinfo {author} {\bibfnamefont {R.~R.}\ \bibnamefont {Netz}},\
  }\bibfield  {title} {\bibinfo {title} {Pumping fluids with periodically
  beating grafted elastic filaments},\ }\href@noop {} {\bibfield  {journal}
  {\bibinfo  {journal} {Phys. Rev. Lett.}\ }\textbf {\bibinfo {volume} {96}},\
  \bibinfo {pages} {158101} (\bibinfo {year} {2006})}\BibitemShut {NoStop}%
\bibitem [{\citenamefont {Zheng}\ \emph {et~al.}(2023)\citenamefont {Zheng},
  \citenamefont {Brandenbourger}, \citenamefont {Robinet}, \citenamefont
  {Schall}, \citenamefont {Lerner},\ and\ \citenamefont
  {Coulais}}]{zheng2023self}%
  \BibitemOpen
  \bibfield  {author} {\bibinfo {author} {\bibfnamefont {E.}~\bibnamefont
  {Zheng}}, \bibinfo {author} {\bibfnamefont {M.}~\bibnamefont
  {Brandenbourger}}, \bibinfo {author} {\bibfnamefont {L.}~\bibnamefont
  {Robinet}}, \bibinfo {author} {\bibfnamefont {P.}~\bibnamefont {Schall}},
  \bibinfo {author} {\bibfnamefont {E.}~\bibnamefont {Lerner}},\ and\ \bibinfo
  {author} {\bibfnamefont {C.}~\bibnamefont {Coulais}},\ }\bibfield  {title}
  {\bibinfo {title} {Self-oscillation and synchronization transitions in
  elastoactive structures},\ }\href
  {https://doi.org/10.1103/PhysRevLett.130.178202} {\bibfield  {journal}
  {\bibinfo  {journal} {Phys. Rev. Lett.}\ }\textbf {\bibinfo {volume} {130}},\
  \bibinfo {pages} {178202} (\bibinfo {year} {2023})}\BibitemShut {NoStop}%
\bibitem [{\citenamefont {Yuan}\ \emph {et~al.}(2014)\citenamefont {Yuan},
  \citenamefont {Raizen},\ and\ \citenamefont {Bau}}]{yuan2014gait}%
  \BibitemOpen
  \bibfield  {author} {\bibinfo {author} {\bibfnamefont {J.}~\bibnamefont
  {Yuan}}, \bibinfo {author} {\bibfnamefont {D.~M.}\ \bibnamefont {Raizen}},\
  and\ \bibinfo {author} {\bibfnamefont {H.~H.}\ \bibnamefont {Bau}},\
  }\bibfield  {title} {\bibinfo {title} {{Gait synchronization in
  Caenorhabditis elegans}},\ }\href@noop {} {\bibfield  {journal} {\bibinfo
  {journal} {Proc. Natl. Acad. Sci. U. S. A.}\ }\textbf {\bibinfo {volume}
  {111}},\ \bibinfo {pages} {6865} (\bibinfo {year} {2014})}\BibitemShut
  {NoStop}%
\bibitem [{\citenamefont {Chelakkot}\ \emph {et~al.}(2021)\citenamefont
  {Chelakkot}, \citenamefont {Hagan},\ and\ \citenamefont
  {Gopinath}}]{chelakkot2021synchronized}%
  \BibitemOpen
  \bibfield  {author} {\bibinfo {author} {\bibfnamefont {R.}~\bibnamefont
  {Chelakkot}}, \bibinfo {author} {\bibfnamefont {M.~F.}\ \bibnamefont
  {Hagan}},\ and\ \bibinfo {author} {\bibfnamefont {A.}~\bibnamefont
  {Gopinath}},\ }\bibfield  {title} {\bibinfo {title} {Synchronized
  oscillations, traveling waves, and jammed clusters induced by steric
  interactions in active filament arrays},\ }\href@noop {} {\bibfield
  {journal} {\bibinfo  {journal} {Soft Matter}\ }\textbf {\bibinfo {volume}
  {17}},\ \bibinfo {pages} {1091} (\bibinfo {year} {2021})}\BibitemShut
  {NoStop}%
\bibitem [{\citenamefont {Guirao}\ and\ \citenamefont
  {Joanny}(2007)}]{guirao2007spontaneous}%
  \BibitemOpen
  \bibfield  {author} {\bibinfo {author} {\bibfnamefont {B.}~\bibnamefont
  {Guirao}}\ and\ \bibinfo {author} {\bibfnamefont {J.-F.}\ \bibnamefont
  {Joanny}},\ }\bibfield  {title} {\bibinfo {title} {Spontaneous creation of
  macroscopic flow and metachronal waves in an array of cilia},\ }\href@noop {}
  {\bibfield  {journal} {\bibinfo  {journal} {Biophys. J.}\ }\textbf {\bibinfo
  {volume} {92}},\ \bibinfo {pages} {1900} (\bibinfo {year}
  {2007})}\BibitemShut {NoStop}%
\bibitem [{\citenamefont {Peshkov}\ \emph {et~al.}(2022)\citenamefont
  {Peshkov}, \citenamefont {McGaffigan},\ and\ \citenamefont
  {Quillen}}]{peshkov2022synchronized}%
  \BibitemOpen
  \bibfield  {author} {\bibinfo {author} {\bibfnamefont {A.}~\bibnamefont
  {Peshkov}}, \bibinfo {author} {\bibfnamefont {S.}~\bibnamefont
  {McGaffigan}},\ and\ \bibinfo {author} {\bibfnamefont {A.~C.}\ \bibnamefont
  {Quillen}},\ }\bibfield  {title} {\bibinfo {title} {{Synchronized
  oscillations in swarms of nematode \emph{Turbatrix aceti}}},\ }\href@noop {}
  {\bibfield  {journal} {\bibinfo  {journal} {Soft Matter}\ }\textbf {\bibinfo
  {volume} {18}},\ \bibinfo {pages} {1174} (\bibinfo {year}
  {2022})}\BibitemShut {NoStop}%
\bibitem [{\citenamefont {Quillen}\ \emph {et~al.}(2021)\citenamefont
  {Quillen}, \citenamefont {Peshkov}, \citenamefont {Wright},\ and\
  \citenamefont {McGaffigan}}]{quillen2021metachronal}%
  \BibitemOpen
  \bibfield  {author} {\bibinfo {author} {\bibfnamefont {A.}~\bibnamefont
  {Quillen}}, \bibinfo {author} {\bibfnamefont {A.}~\bibnamefont {Peshkov}},
  \bibinfo {author} {\bibfnamefont {E.}~\bibnamefont {Wright}},\ and\ \bibinfo
  {author} {\bibfnamefont {S.}~\bibnamefont {McGaffigan}},\ }\bibfield  {title}
  {\bibinfo {title} {{Metachronal waves in concentrations of swimming
  \emph{Turbatrix aceti} nematodes and an oscillator chain model for their
  coordinated motions}},\ }\href@noop {} {\bibfield  {journal} {\bibinfo
  {journal} {Phys. Rev. E.}\ }\textbf {\bibinfo {volume} {104}},\ \bibinfo
  {pages} {014412} (\bibinfo {year} {2021})}\BibitemShut {NoStop}%
\bibitem [{\citenamefont {Goldstein}\ \emph {et~al.}(2016)\citenamefont
  {Goldstein}, \citenamefont {Lauga}, \citenamefont {Pesci},\ and\
  \citenamefont {Proctor}}]{goldstein2016elastohydrodynamic}%
  \BibitemOpen
  \bibfield  {author} {\bibinfo {author} {\bibfnamefont {R.~E.}\ \bibnamefont
  {Goldstein}}, \bibinfo {author} {\bibfnamefont {E.}~\bibnamefont {Lauga}},
  \bibinfo {author} {\bibfnamefont {A.~I.}\ \bibnamefont {Pesci}},\ and\
  \bibinfo {author} {\bibfnamefont {M.~R.}\ \bibnamefont {Proctor}},\
  }\bibfield  {title} {\bibinfo {title} {Elastohydrodynamic synchronization of
  adjacent beating flagella},\ }\href@noop {} {\bibfield  {journal} {\bibinfo
  {journal} {Phys. Rev. Fluids}\ }\textbf {\bibinfo {volume} {1}},\ \bibinfo
  {pages} {073201} (\bibinfo {year} {2016})}\BibitemShut {NoStop}%
\bibitem [{\citenamefont {Heuser}\ \emph {et~al.}(2009)\citenamefont {Heuser},
  \citenamefont {Raytchev}, \citenamefont {Krell}, \citenamefont {Porter},\
  and\ \citenamefont {Nicastro}}]{heuser2009dynein}%
  \BibitemOpen
  \bibfield  {author} {\bibinfo {author} {\bibfnamefont {T.}~\bibnamefont
  {Heuser}}, \bibinfo {author} {\bibfnamefont {M.}~\bibnamefont {Raytchev}},
  \bibinfo {author} {\bibfnamefont {J.}~\bibnamefont {Krell}}, \bibinfo
  {author} {\bibfnamefont {M.~E.}\ \bibnamefont {Porter}},\ and\ \bibinfo
  {author} {\bibfnamefont {D.}~\bibnamefont {Nicastro}},\ }\bibfield  {title}
  {\bibinfo {title} {The dynein regulatory complex is the nexin link and a
  major regulatory node in cilia and flagella},\ }\href@noop {} {\bibfield
  {journal} {\bibinfo  {journal} {J. Cell Biol.}\ }\textbf {\bibinfo {volume}
  {187}},\ \bibinfo {pages} {921} (\bibinfo {year} {2009})}\BibitemShut
  {NoStop}%
\bibitem [{\citenamefont {J{\"u}licher}\ and\ \citenamefont
  {Prost}(1997)}]{julicher1997spontaneous}%
  \BibitemOpen
  \bibfield  {author} {\bibinfo {author} {\bibfnamefont {F.}~\bibnamefont
  {J{\"u}licher}}\ and\ \bibinfo {author} {\bibfnamefont {J.}~\bibnamefont
  {Prost}},\ }\bibfield  {title} {\bibinfo {title} {Spontaneous oscillations of
  collective molecular motors},\ }\href@noop {} {\bibfield  {journal} {\bibinfo
   {journal} {Phys. Rev. Lett.}\ }\textbf {\bibinfo {volume} {78}},\ \bibinfo
  {pages} {4510} (\bibinfo {year} {1997})}\BibitemShut {NoStop}%
\bibitem [{\citenamefont {Gray}\ and\ \citenamefont
  {Hancock}(1955)}]{gray1955propulsion}%
  \BibitemOpen
  \bibfield  {author} {\bibinfo {author} {\bibfnamefont {J.}~\bibnamefont
  {Gray}}\ and\ \bibinfo {author} {\bibfnamefont {G.}~\bibnamefont {Hancock}},\
  }\bibfield  {title} {\bibinfo {title} {The propulsion of sea-urchin
  spermatozoa},\ }\href@noop {} {\bibfield  {journal} {\bibinfo  {journal} {J.
  Exp. Biol.}\ }\textbf {\bibinfo {volume} {32}},\ \bibinfo {pages} {802}
  (\bibinfo {year} {1955})}\BibitemShut {NoStop}%
\bibitem [{\citenamefont {Rotne}\ and\ \citenamefont
  {Prager}(1969)}]{rotne1969variational}%
  \BibitemOpen
  \bibfield  {author} {\bibinfo {author} {\bibfnamefont {J.}~\bibnamefont
  {Rotne}}\ and\ \bibinfo {author} {\bibfnamefont {S.}~\bibnamefont {Prager}},\
  }\bibfield  {title} {\bibinfo {title} {Variational treatment of hydrodynamic
  interaction in polymers},\ }\href@noop {} {\bibfield  {journal} {\bibinfo
  {journal} {J. Chem. Phys.}\ }\textbf {\bibinfo {volume} {50}},\ \bibinfo
  {pages} {4831} (\bibinfo {year} {1969})}\BibitemShut {NoStop}%
\bibitem [{\citenamefont {Yamakawa}(1970)}]{yamakawa1970transport}%
  \BibitemOpen
  \bibfield  {author} {\bibinfo {author} {\bibfnamefont {H.}~\bibnamefont
  {Yamakawa}},\ }\bibfield  {title} {\bibinfo {title} {Transport properties of
  polymer chains in dilute solution: hydrodynamic interaction},\ }\href@noop {}
  {\bibfield  {journal} {\bibinfo  {journal} {J. Chem. Phys.}\ }\textbf
  {\bibinfo {volume} {53}},\ \bibinfo {pages} {436} (\bibinfo {year}
  {1970})}\BibitemShut {NoStop}%
\bibitem [{\citenamefont {Tornberg}\ and\ \citenamefont
  {Shelley}(2004)}]{tornberg2004simulating}%
  \BibitemOpen
  \bibfield  {author} {\bibinfo {author} {\bibfnamefont {A.-K.}\ \bibnamefont
  {Tornberg}}\ and\ \bibinfo {author} {\bibfnamefont {M.~J.}\ \bibnamefont
  {Shelley}},\ }\bibfield  {title} {\bibinfo {title} {Simulating the dynamics
  and interactions of flexible fibers in stokes flows},\ }\href@noop {}
  {\bibfield  {journal} {\bibinfo  {journal} {J. Comp. Phys.}\ }\textbf
  {\bibinfo {volume} {196}},\ \bibinfo {pages} {8} (\bibinfo {year}
  {2004})}\BibitemShut {NoStop}%
\bibitem [{\citenamefont {Lauga}\ and\ \citenamefont
  {Powers}(2009)}]{lauga2009hydrodynamics}%
  \BibitemOpen
  \bibfield  {author} {\bibinfo {author} {\bibfnamefont {E.}~\bibnamefont
  {Lauga}}\ and\ \bibinfo {author} {\bibfnamefont {T.~R.}\ \bibnamefont
  {Powers}},\ }\bibfield  {title} {\bibinfo {title} {The hydrodynamics of
  swimming microorganisms},\ }\href@noop {} {\bibfield  {journal} {\bibinfo
  {journal} {Rep. Prog. Phys.}\ }\textbf {\bibinfo {volume} {72}},\ \bibinfo
  {pages} {096601} (\bibinfo {year} {2009})}\BibitemShut {NoStop}%
\bibitem [{\citenamefont {Wiggins}\ and\ \citenamefont
  {Goldstein}(1998)}]{wiggins1998flexive}%
  \BibitemOpen
  \bibfield  {author} {\bibinfo {author} {\bibfnamefont {C.~H.}\ \bibnamefont
  {Wiggins}}\ and\ \bibinfo {author} {\bibfnamefont {R.~E.}\ \bibnamefont
  {Goldstein}},\ }\bibfield  {title} {\bibinfo {title} {{Flexive and propulsive
  dynamics of elastica at low Reynolds number}},\ }\href@noop {} {\bibfield
  {journal} {\bibinfo  {journal} {Phys. Rev. Lett.}\ }\textbf {\bibinfo
  {volume} {80}},\ \bibinfo {pages} {3879} (\bibinfo {year}
  {1998})}\BibitemShut {NoStop}%
\bibitem [{\citenamefont {Nakao}(2016)}]{nakao2016phase}%
  \BibitemOpen
  \bibfield  {author} {\bibinfo {author} {\bibfnamefont {H.}~\bibnamefont
  {Nakao}},\ }\bibfield  {title} {\bibinfo {title} {Phase reduction approach to
  synchronisation of nonlinear oscillators},\ }\href@noop {} {\bibfield
  {journal} {\bibinfo  {journal} {Contemporary Physics}\ }\textbf {\bibinfo
  {volume} {57}},\ \bibinfo {pages} {188} (\bibinfo {year} {2016})}\BibitemShut
  {NoStop}%
\bibitem [{\citenamefont {Kawamura}\ and\ \citenamefont
  {Tsubaki}(2018)}]{kawamura2018phase}%
  \BibitemOpen
  \bibfield  {author} {\bibinfo {author} {\bibfnamefont {Y.}~\bibnamefont
  {Kawamura}}\ and\ \bibinfo {author} {\bibfnamefont {R.}~\bibnamefont
  {Tsubaki}},\ }\bibfield  {title} {\bibinfo {title} {Phase reduction approach
  to elastohydrodynamic synchronization of beating flagella},\ }\href@noop {}
  {\bibfield  {journal} {\bibinfo  {journal} {Phys. Rev. E.}\ }\textbf
  {\bibinfo {volume} {97}},\ \bibinfo {pages} {022212} (\bibinfo {year}
  {2018})}\BibitemShut {NoStop}%
\bibitem [{\citenamefont {Man}\ and\ \citenamefont
  {Kanso}(2020)}]{man2020multisynchrony}%
  \BibitemOpen
  \bibfield  {author} {\bibinfo {author} {\bibfnamefont {Y.}~\bibnamefont
  {Man}}\ and\ \bibinfo {author} {\bibfnamefont {E.}~\bibnamefont {Kanso}},\
  }\bibfield  {title} {\bibinfo {title} {Multisynchrony in active
  microfilaments},\ }\href@noop {} {\bibfield  {journal} {\bibinfo  {journal}
  {Phys. Rev. Lett.}\ }\textbf {\bibinfo {volume} {125}},\ \bibinfo {pages}
  {148101} (\bibinfo {year} {2020})}\BibitemShut {NoStop}%
\bibitem [{\citenamefont {Quillen}(2023)}]{quillen2023robust}%
  \BibitemOpen
  \bibfield  {author} {\bibinfo {author} {\bibfnamefont {A.}~\bibnamefont
  {Quillen}},\ }\bibfield  {title} {\bibinfo {title} {Robust formation of
  metachronal waves in directional chains of phase oscillators},\ }\href@noop
  {} {\bibfield  {journal} {\bibinfo  {journal} {Phys. Rev. E}\ }\textbf
  {\bibinfo {volume} {107}},\ \bibinfo {pages} {034401} (\bibinfo {year}
  {2023})}\BibitemShut {NoStop}%
\bibitem [{\citenamefont {Guo}\ \emph {et~al.}(2018)\citenamefont {Guo},
  \citenamefont {Fauci}, \citenamefont {Shelley},\ and\ \citenamefont
  {Kanso}}]{guo2018bistability}%
  \BibitemOpen
  \bibfield  {author} {\bibinfo {author} {\bibfnamefont {H.}~\bibnamefont
  {Guo}}, \bibinfo {author} {\bibfnamefont {L.}~\bibnamefont {Fauci}}, \bibinfo
  {author} {\bibfnamefont {M.}~\bibnamefont {Shelley}},\ and\ \bibinfo {author}
  {\bibfnamefont {E.}~\bibnamefont {Kanso}},\ }\bibfield  {title} {\bibinfo
  {title} {Bistability in the synchronization of actuated microfilaments},\
  }\href@noop {} {\bibfield  {journal} {\bibinfo  {journal} {J. Fluid Mech.}\
  }\textbf {\bibinfo {volume} {836}},\ \bibinfo {pages} {304} (\bibinfo {year}
  {2018})}\BibitemShut {NoStop}%
\bibitem [{\citenamefont {Young}(2009)}]{young2009hydrodynamic}%
  \BibitemOpen
  \bibfield  {author} {\bibinfo {author} {\bibfnamefont {Y.-N.}\ \bibnamefont
  {Young}},\ }\bibfield  {title} {\bibinfo {title} {Hydrodynamic interactions
  between two semiflexible inextensible filaments in stokes flow},\ }\href@noop
  {} {\bibfield  {journal} {\bibinfo  {journal} {Phys. Rev. E.}\ }\textbf
  {\bibinfo {volume} {79}},\ \bibinfo {pages} {046317} (\bibinfo {year}
  {2009})}\BibitemShut {NoStop}%
\bibitem [{\citenamefont {Vilfan}\ and\ \citenamefont
  {J{\"u}licher}(2006)}]{vilfan2006hydrodynamic}%
  \BibitemOpen
  \bibfield  {author} {\bibinfo {author} {\bibfnamefont {A.}~\bibnamefont
  {Vilfan}}\ and\ \bibinfo {author} {\bibfnamefont {F.}~\bibnamefont
  {J{\"u}licher}},\ }\bibfield  {title} {\bibinfo {title} {Hydrodynamic flow
  patterns and synchronization of beating cilia},\ }\href@noop {} {\bibfield
  {journal} {\bibinfo  {journal} {Phys. Rev. Lett.}\ }\textbf {\bibinfo
  {volume} {96}},\ \bibinfo {pages} {058102} (\bibinfo {year}
  {2006})}\BibitemShut {NoStop}%
\bibitem [{\citenamefont {Lighthill}(1975)}]{lighthill1975mathematical}%
  \BibitemOpen
  \bibfield  {author} {\bibinfo {author} {\bibfnamefont {S.~J.}\ \bibnamefont
  {Lighthill}},\ }\href@noop {} {\emph {\bibinfo {title} {Mathematical biofluid
  dynamics}}}\ (\bibinfo  {publisher} {SIAM},\ \bibinfo {year}
  {1975})\BibitemShut {NoStop}%
\bibitem [{\citenamefont {Chakrabarti}\ and\ \citenamefont
  {Saintillan}(2019)}]{chakrabarti2019spontaneous}%
  \BibitemOpen
  \bibfield  {author} {\bibinfo {author} {\bibfnamefont {B.}~\bibnamefont
  {Chakrabarti}}\ and\ \bibinfo {author} {\bibfnamefont {D.}~\bibnamefont
  {Saintillan}},\ }\bibfield  {title} {\bibinfo {title} {Spontaneous
  oscillations, beating patterns, and hydrodynamics of active microfilaments},\
  }\href@noop {} {\bibfield  {journal} {\bibinfo  {journal} {Phys. Rev.
  Fluids}\ }\textbf {\bibinfo {volume} {4}},\ \bibinfo {pages} {043102}
  (\bibinfo {year} {2019})}\BibitemShut {NoStop}%
\bibitem [{\citenamefont {Fang-Yen}\ \emph {et~al.}(2010)\citenamefont
  {Fang-Yen}, \citenamefont {Wyart}, \citenamefont {Xie}, \citenamefont
  {Kawai}, \citenamefont {Kodger}, \citenamefont {Chen}, \citenamefont {Wen},\
  and\ \citenamefont {Samuel}}]{fang2010biomechanical}%
  \BibitemOpen
  \bibfield  {author} {\bibinfo {author} {\bibfnamefont {C.}~\bibnamefont
  {Fang-Yen}}, \bibinfo {author} {\bibfnamefont {M.}~\bibnamefont {Wyart}},
  \bibinfo {author} {\bibfnamefont {J.}~\bibnamefont {Xie}}, \bibinfo {author}
  {\bibfnamefont {R.}~\bibnamefont {Kawai}}, \bibinfo {author} {\bibfnamefont
  {T.}~\bibnamefont {Kodger}}, \bibinfo {author} {\bibfnamefont
  {S.}~\bibnamefont {Chen}}, \bibinfo {author} {\bibfnamefont {Q.}~\bibnamefont
  {Wen}},\ and\ \bibinfo {author} {\bibfnamefont {A.~D.}\ \bibnamefont
  {Samuel}},\ }\bibfield  {title} {\bibinfo {title} {{Biomechanical analysis of
  gait adaptation in the nematode Caenorhabditis elegans}},\ }\href@noop {}
  {\bibfield  {journal} {\bibinfo  {journal} {Proc. Natl. Acad. Sci. U. S. A.}\
  }\textbf {\bibinfo {volume} {107}},\ \bibinfo {pages} {20323} (\bibinfo
  {year} {2010})}\BibitemShut {NoStop}%
\bibitem [{\citenamefont {Sznitman}\ \emph {et~al.}(2010)\citenamefont
  {Sznitman}, \citenamefont {Purohit}, \citenamefont {Krajacic}, \citenamefont
  {Lamitina},\ and\ \citenamefont {Arratia}}]{sznitman2010material}%
  \BibitemOpen
  \bibfield  {author} {\bibinfo {author} {\bibfnamefont {J.}~\bibnamefont
  {Sznitman}}, \bibinfo {author} {\bibfnamefont {P.~K.}\ \bibnamefont
  {Purohit}}, \bibinfo {author} {\bibfnamefont {P.}~\bibnamefont {Krajacic}},
  \bibinfo {author} {\bibfnamefont {T.}~\bibnamefont {Lamitina}},\ and\
  \bibinfo {author} {\bibfnamefont {P.~E.}\ \bibnamefont {Arratia}},\
  }\bibfield  {title} {\bibinfo {title} {{Material properties of Caenorhabditis
  elegans swimming at low Reynolds number}},\ }\href@noop {} {\bibfield
  {journal} {\bibinfo  {journal} {Biophys. J.}\ }\textbf {\bibinfo {volume}
  {98}},\ \bibinfo {pages} {617} (\bibinfo {year} {2010})}\BibitemShut
  {NoStop}%
\bibitem [{\citenamefont {Xu}\ \emph {et~al.}(2016)\citenamefont {Xu},
  \citenamefont {Wilson}, \citenamefont {Okamoto}, \citenamefont {Shao},
  \citenamefont {Dutcher},\ and\ \citenamefont {Bayly}}]{xu2016flexural}%
  \BibitemOpen
  \bibfield  {author} {\bibinfo {author} {\bibfnamefont {G.}~\bibnamefont
  {Xu}}, \bibinfo {author} {\bibfnamefont {K.~S.}\ \bibnamefont {Wilson}},
  \bibinfo {author} {\bibfnamefont {R.~J.}\ \bibnamefont {Okamoto}}, \bibinfo
  {author} {\bibfnamefont {J.-Y.}\ \bibnamefont {Shao}}, \bibinfo {author}
  {\bibfnamefont {S.~K.}\ \bibnamefont {Dutcher}},\ and\ \bibinfo {author}
  {\bibfnamefont {P.~V.}\ \bibnamefont {Bayly}},\ }\bibfield  {title} {\bibinfo
  {title} {Flexural rigidity and shear stiffness of flagella estimated from
  induced bends and counterbends},\ }\href@noop {} {\bibfield  {journal}
  {\bibinfo  {journal} {Biophys. J.}\ }\textbf {\bibinfo {volume} {110}},\
  \bibinfo {pages} {2759} (\bibinfo {year} {2016})}\BibitemShut {NoStop}%
\bibitem [{\citenamefont {Gilpin}\ \emph {et~al.}(2015)\citenamefont {Gilpin},
  \citenamefont {Uppaluri},\ and\ \citenamefont
  {Brangwynne}}]{gilpin2015worms}%
  \BibitemOpen
  \bibfield  {author} {\bibinfo {author} {\bibfnamefont {W.}~\bibnamefont
  {Gilpin}}, \bibinfo {author} {\bibfnamefont {S.}~\bibnamefont {Uppaluri}},\
  and\ \bibinfo {author} {\bibfnamefont {C.~P.}\ \bibnamefont {Brangwynne}},\
  }\bibfield  {title} {\bibinfo {title} {{Worms under pressure: bulk mechanical
  properties of C. elegans are independent of the cuticle}},\ }\href@noop {}
  {\bibfield  {journal} {\bibinfo  {journal} {Biophys. J.}\ }\textbf {\bibinfo
  {volume} {108}},\ \bibinfo {pages} {1887} (\bibinfo {year}
  {2015})}\BibitemShut {NoStop}%
\bibitem [{\citenamefont {Elfring}\ and\ \citenamefont
  {Lauga}(2009)}]{elfring2009hydrodynamic}%
  \BibitemOpen
  \bibfield  {author} {\bibinfo {author} {\bibfnamefont {G.~J.}\ \bibnamefont
  {Elfring}}\ and\ \bibinfo {author} {\bibfnamefont {E.}~\bibnamefont
  {Lauga}},\ }\bibfield  {title} {\bibinfo {title} {Hydrodynamic phase locking
  of swimming microorganisms},\ }\href@noop {} {\bibfield  {journal} {\bibinfo
  {journal} {Phys. Rev. Lett.}\ }\textbf {\bibinfo {volume} {103}},\ \bibinfo
  {pages} {088101} (\bibinfo {year} {2009})}\BibitemShut {NoStop}%
\bibitem [{\citenamefont {Leptos}\ \emph {et~al.}(2013)\citenamefont {Leptos},
  \citenamefont {Wan}, \citenamefont {Polin}, \citenamefont {Tuval},
  \citenamefont {Pesci},\ and\ \citenamefont
  {Goldstein}}]{leptos2013antiphase}%
  \BibitemOpen
  \bibfield  {author} {\bibinfo {author} {\bibfnamefont {K.~C.}\ \bibnamefont
  {Leptos}}, \bibinfo {author} {\bibfnamefont {K.~Y.}\ \bibnamefont {Wan}},
  \bibinfo {author} {\bibfnamefont {M.}~\bibnamefont {Polin}}, \bibinfo
  {author} {\bibfnamefont {I.}~\bibnamefont {Tuval}}, \bibinfo {author}
  {\bibfnamefont {A.~I.}\ \bibnamefont {Pesci}},\ and\ \bibinfo {author}
  {\bibfnamefont {R.~E.}\ \bibnamefont {Goldstein}},\ }\bibfield  {title}
  {\bibinfo {title} {Antiphase synchronization in a flagellar-dominance mutant
  of chlamydomonas},\ }\href@noop {} {\bibfield  {journal} {\bibinfo  {journal}
  {Phys. Rev. Lett.}\ }\textbf {\bibinfo {volume} {111}},\ \bibinfo {pages}
  {158101} (\bibinfo {year} {2013})}\BibitemShut {NoStop}%
\end{thebibliography}

%

\clearpage 

\section{Supplemental Materials}

\setcounter{equation}{0}
\setcounter{figure}{0}
\setcounter{table}{0}
\renewcommand{\theequation}{S\arabic{equation}}
\renewcommand{\thefigure}{S\arabic{figure}}
\renewcommand{\thetable}{S\arabic{table}}

\section{Supplemental Movies}
\begin{itemize}
\item SM Movie 1: Beat dynamics of a single filament at $\theta=0^\circ$. 
\item SM Movie 2: Dynamic of two interacting filaments at $\theta=0^\circ$ with varying sizes of mismatch $\Delta l$. 
\item SM Movie 3: Dynamics of $N=10$ filaments at 
$\theta=30^\circ$, $45^\circ$, $60^\circ$, and $72^\circ$.   
\item SM Movie 4: Demonstration of the effect of SI on two-filament dynamics starting from a V-shape configuration. 
\item SM Movie 5: Many filament dynamics at $\theta=0^\circ$ with varying $N$, without periodic boundary condition. 
\item SM Movie 6: Many filament dynamics at $\theta=60^\circ$ with varying $N$, without periodic boundary condition. 
\end{itemize}

\section{Elastohydrodynamic model}
We consider $N$ filaments, each with a length of $L$ and a cross-sectional radius of $a$, tilted with $\theta$ while maintaining a body-to-body separation distance $d$ within a two-dimensional plane (Fig.~1).
The head of each filament is hinged along the $y'$-axis. The head-to-head distance depends on the tilt angle $\theta$ and is separated by $d_{h}=d/\cos{\theta}$ for the given separation $d$. 
An aspect ratio of the filament is $L/(2a)=50$ and the scale of $d$ used in the model is in the range of $a\ll d\ll L$. 
Under an assumption of small undulation compared to the body length, the curvilinear coordinate along the filament ($0\leq s\leq L$) is approximated to $s\approx x$. 

At low Reynolds number, the force balance on the $i$th filament should be established among viscous drag ($f^i_{vis}$), bending ($f^i_{ben}$), active forces ($f^i_{act}$), hydrodynamic couplings ($\sum\limits_{j=i-1,i+1}f^{j\rightarrow i}_{hyd}$) and steric interactions ($\sum\limits_{j=i-1,i+1}{f^{j\rightarrow i}_{rep}}$) from the neighboring filaments, such that the sum of all the force components should vanish, $\sum_{\alpha}f^i_{\alpha}=0$.   

Each force term can be related to the waveform of filaments $h^i(x,t)$. 
The force associated with viscous drag is given as  
\begin{align}
    f^i_{vis}=-\zeta_{\perp}\partial_t h^i
\end{align}
where $\zeta_{\perp}\approx 4\pi\mu/(\ln{L/a})\approx \pi \mu$ is the drag coefficient for a slender body \cite{lighthill1975mathematical,lauga2009hydrodynamics}. 
The force arising from the bending penalty can be obtained from the functional derivative of the bending energy $(\kappa/2)\int_0^L dx (\partial^2_x h^i)^2$, 
\begin{align}
    f^i_{ben} = -\kappa \partial_x^4 h^i
    \end{align}
where $\kappa$ is the coefficient of bending stiffness. 

For the active force we adopt the phenomenological model by Goldstein \textit{et. al} \cite{goldstein2016elastohydrodynamic}
\begin{align}
    f^i_{act} = -A\partial_x h^i -B\partial^2_x h^i+C\left(\partial^2_x h^i\right)^3
\end{align}
where the first term ($A>0$) generates propagating waves along the centerline of filament ($x$-axis in Fig.~1). 
The second and the third term break head-tail symmetry of the amplitude of transversal undulations, which is magnified by the second term ($B>0$) and suppressed by the third term ($C>0$), saturates to a finite value. 

The hydrodynamic force from $j$th to $i$th filaments is defined via the fundamental singular solutions of Stokes equations: 
\begin{align}
    f^{j\rightarrow i}_{hyd} (x^i,t)/\zeta_{\perp}=  \textbf{e}_y\cdot \int_0^L dx^j \left(\textbf{G}(r)\cdot \textbf{f}^j_{vis}(x^j,t)\right)
    \label{eqn:hyd}
\end{align}
where $\textbf{e}_y=(0,1)$, $ \textbf{G}(r)$ is the sum of Stokeslet and doublet  
\begin{align}
    \textbf{G}(r)=\frac{1}{8\pi\mu r}\left[ \left(\textbf{I}+\frac{\textbf{r}\textbf{r}}{r^2}\right)+\frac{a^2}{2r^2}\left(\textbf{I}-3\frac{\textbf{r}\textbf{r}}{r^2}\right)\right]
    \label{eqn:Stokeslet}
\end{align}
with $\textbf{r}$ denoting the relative position vector from $(x^j,h^j)$ to $(x^i,h^i)$ and $r=|\textbf{r}|$. 
The second term, corresponding to the doublet, corrects the boundary condition when the velocity is computed on the surface of the slender body near the singularity distribution. In Eq.\ref{eqn:hyd}, the term associated with viscous force of $j$-th filament, $\textbf{f}^j_{vis}(x^j,t)$, is also obtained from the resistive force theory,  
$f^j_{vis}(x^j,t)=\textbf{e}_y\cdot\textbf{f}^j_{vis}(x^j,t) = \zeta_{\perp}\partial_t h^j(x^j,t)$. 
It should be emphasized that for a non-zero tilt ($\Delta l\neq0$), 
the nearest point of adjacent chain to $x^i$ is $x^j=x^i\pm \Delta l$ for $j=i\mp1$. 
Taken together, 
Eq.\ref{eqn:hyd} is simplified to the following form~\cite{goldstein2016elastohydrodynamic}:
\begin{align}
   f^{j\rightarrow i}_{hyd} (x^i,t)/\zeta_{\perp}= \epsilon_h \partial_t h^j(x^j,t). 
\label{eqn:hydro}
\end{align}

For the steric repulsion between the adjacent filaments, we use the Weeks-Chandler-Anderson (WCA) potential 
\begin{align}
U_R(r)=
\begin{cases} 4\epsilon_r\left[\left(\frac{\sigma}{r}\right)^{12}-\left(\frac{\sigma}{r}\right)^6\right]+\epsilon_r & \text{for } r\leq 2^{1/6}\sigma \\ 0 & \text{otherwise} \end{cases}
\end{align}
where $\sigma$ is a length scale at which $U_R(\sigma)=0$ and $\epsilon_r$ is a scale of interaction energy. 
The repulsive force due to the steric interaction is obtained from 
\begin{align}
    f^{j\rightarrow i}_{rep} (x^i,t)= {\textbf{e}_y} \cdot\int_0^L dx^j \left(-\frac{\partial U_R(r)}{\partial r}\right)\frac{\textbf{r}}{r}. 
\end{align}

For a single active elastic filament demonstrating a beating dynamics, 
two dimensionless parameters can characterize its dynamics: (i) activity number, the active force relative to the bending force; (ii) sperm number, the beat frequency  relative to the rate of the bending wave relaxation in a viscous fluid  \cite{wiggins1998flexive,chakrabarti2019spontaneous}. 
A waveform generated from a collective dynamics of filaments in a dense suspension is also characterized by the two parameters of a single filament \cite{yang2008cooperation}.   
In our system, the activity and sperm numbers are given as
$\mu_a \equiv \frac{AL}{\kappa/(2aL)}=160$, $Sp \equiv \left(L^4\zeta_{\perp}/\kappa\tau_a\right)^{1/4} = \left(L^4 \tau_b /l_s^4\tau_a\right)^{1/4}=L/l_s=20$.
Here, we assume that the characteristic time scale of active motion, $\tau_a$, is the same as that of bending relaxation time scale, $\tau_b$, and use it as a characteristic time scale of the system, $\tau_s=\tau_b=\tau_a$. 

The following expression that considers the WCA potential is numerically computed:
\begin{widetext}
\begin{align}
f^{j\rightarrow i}_{rep} (x^i,t)/\zeta_{\perp} = \begin{cases} (4\epsilon_r/\zeta_{\perp}) \int_0^L \left(\frac{12\sigma^{12}}{r^{14}}-\frac{6\sigma^6}{r^8}\right)[h^i-h^j+(i-j)d+2a]dx^j & \text{if } r\leq 2^{1/6}\sigma, \\ 0 & \text{otherwise}. \end{cases}
\label{eqn:si}
\end{align}
for the steric repulsion between the neighboring filaments,
and 
\begin{align}
    f^{{\rm w}\rightarrow i}_{rep} (x^i,t)/\zeta_{\perp}&= \zeta_{\perp}^{-1}{\textbf{e}_y} \cdot\int_0^{L_{\rm w}} dx^{\rm w} \left(-\frac{\partial U_R(r_{\rm w})}{\partial r_{\rm w}}\right)\frac{\textbf{r}_{\rm w}}{r_{\rm w}}\nonumber\\
&= \begin{cases} (4\epsilon_r/\zeta_{\perp}) \int_0^{L_{\rm w}}\left(\frac{12\sigma^{12}}{r_{\rm w}^{14}}-\frac{6\sigma^6}{r_{\rm w}^8}\right)[y^i-y^{\rm w}]dx^{\rm w} & \text{if } r_{\rm w}\leq 2^{1/6}\sigma, \\ 0 & \text{otherwise}. \end{cases}
\label{eqn:wi}
\end{align}
for the wall repulsion, 
where $\bold{r}_{\rm w}=(x^i,y^i)-(x^{\rm w},y^{\rm w})$, $||\bold{r}_{\rm w}||=r_{\rm w}=\sqrt{(x^i-x^{\rm w})^2+(y^i-y^{\rm w})^2}$, $\bold{e}_y\cdot\bold{r}_{\rm w}=y^i-y^{\rm w}$, and 
$\left(x^i-x^{\rm w},y^i-y^{\rm w}\right)=\left(x^i+(i-1)\Delta l-x^{\rm w}\sin\theta+d_{\rm w}\cos\theta, h^i(x^i,t)+(i-1)d-x^{\rm w}\cos\theta-d_{\rm w}\sin\theta\right)$.
\end{widetext}
The boundary conditions were imposed to create a hinged head ($h^i(0,t)=h^i_{xx}(0,t)=0$) and a free tail ($h^i_{xx}(L,t)=0$ and $h^i_{xxx}(L,t)=0$). 
As initial conditions, we used $N$ straight filaments in a parallel alignment, and studied the dynamics for various $N$. 
The governing equation (Eq.\ref{eqn:main}) was integrated using the finite difference and the semi-implicit scheme \cite{tornberg2004simulating}. 

Using the relations, $(\tau_s\kappa/\zeta_{\perp}l_s^4)=1$, $(\tau_sA/\zeta_{\perp}l_s)=c$, 
$(\tau_sB/\zeta_{\perp}l_s^2)=2$,  and $(\tau_sC/\zeta_{\perp}l_s^4)=1$, which determine  
the characteristic length and time scales of the system as $l_s=\left(2\kappa/B\right)^{1/2}$
and $\tau_s=(4\kappa\zeta_\perp/B^2)$, 
we non-dimensionalize the force balance equation for each filament as 
\begin{widetext}
\begin{align}
 \frac{\partial \hat{h}^i}{\partial \hat{t}}=\left[-c\frac{\partial \hat{h}^i}{\partial \hat{x}} - 2  \frac{\partial^2 \hat{h}^i}{\partial \hat{x}^2}+ \left(\frac{\partial^2 \hat{h}^i}{\partial \hat{x}^2}\right)^3 \right] - \frac{\partial^4 \hat{h}^i}{\partial \hat{x}^4} +  \frac{\tau_s}{l_s}\left[\sum_{j=i-1,i+1} \frac{f^{j\rightarrow i}_{hyd}}{ \zeta_{\perp}} + \sum_{j=i-1,i+1}\frac{f^{j\rightarrow i}_{rep}}{\zeta_{\perp}}\right]
 \label{eqn:main}
\end{align}
\end{widetext}
where $\hat{h}=h/l_s$, $\hat{x}=x/l_s$, $\hat{t}=t/\tau_s$, and 
$c = A \kappa^{1/2}(2/B)^{3/2} $ is a dimensionless speed of wave.   
For the parameter values of $\mu=1$ $\text{cP}=10^{-3}$ Pa$\cdot$sec, $B=0.8$ $\mu$N \cite{peshkov2022synchronized}, and $C=\kappa = 10^{-3}$ $\mu$N$\cdot$mm$^2$ (corresponding to the Young's modulus of $E=125$ kPa) \cite{fang2010biomechanical,sznitman2010material, xu2016flexural,gilpin2015worms}, we obtain $l_s=0.05$ mm and 
$\tau_s= 20$ ms. 
When $c=1$, $A = 8$ $\mu$N/mm. We also set the values of $\epsilon_r=A$ $(=5\times10^9 k_BT/l_s^2)$ and $\sigma=2a$. 
In Eq.~1, the hat symbol for non-dimensional length ($\hat{h}$ and $\hat{x}$) and time ($\hat{t}$) is dropped from the expression for simplicity.  

The stability analysis on the linearized elastohydrodynamic equation, 
$\partial_t\hat{h}=-c\partial_x\hat{h}-2\partial_x^2\hat{h}-\partial_x^4\hat{h}$ with $\hat{h}\sim e^{i(\omega t- kx)}$ yields a dispersion relation, 
\begin{align}
i\omega=-ick+2k^2-k^4. 
\end{align}
Thus, for the dynamics of the filament to be oscillatory, 
the condition of $k^2(k^2-2)<0$ imposes  
the wavelength ($k=2\pi/\lambda<\sqrt{2}$) of a single filament always greater than $\sqrt{2}\pi$, 
\begin{align}
\lambda>\sqrt{2}\pi.
\end{align}

\section{Effect of steric interactions on the synchronization of two hydrodynamically coupled filaments}

The effect of SI on the two hydrodynamically coupled filaments are investigated for a system without tilt ($\theta=0^\circ$),  separated by $d=10a$. 
To quantify the process of synchronization between two active filaments ($i,j=1,2$ and $i\neq j$), we calculate a correlation function,  
\begin{align}
C(t)=\frac{1}{L}\int_0^L \frac{1+\left(\partial_x h^1\right)\left(\partial_x h^2\right)}{\sqrt{1+\left(\partial_x h^1\right)^2}\sqrt{1+\left(\partial_x h^2\right)^2}}dx. 
\label{eqn:Ct}
\end{align}
Here, $C(t)=1$ for perfect synchrony and $C(t)=0$ when their dynamics are decoupled.

\begin{figure}[ht!]
	\includegraphics[width=0.9\linewidth]{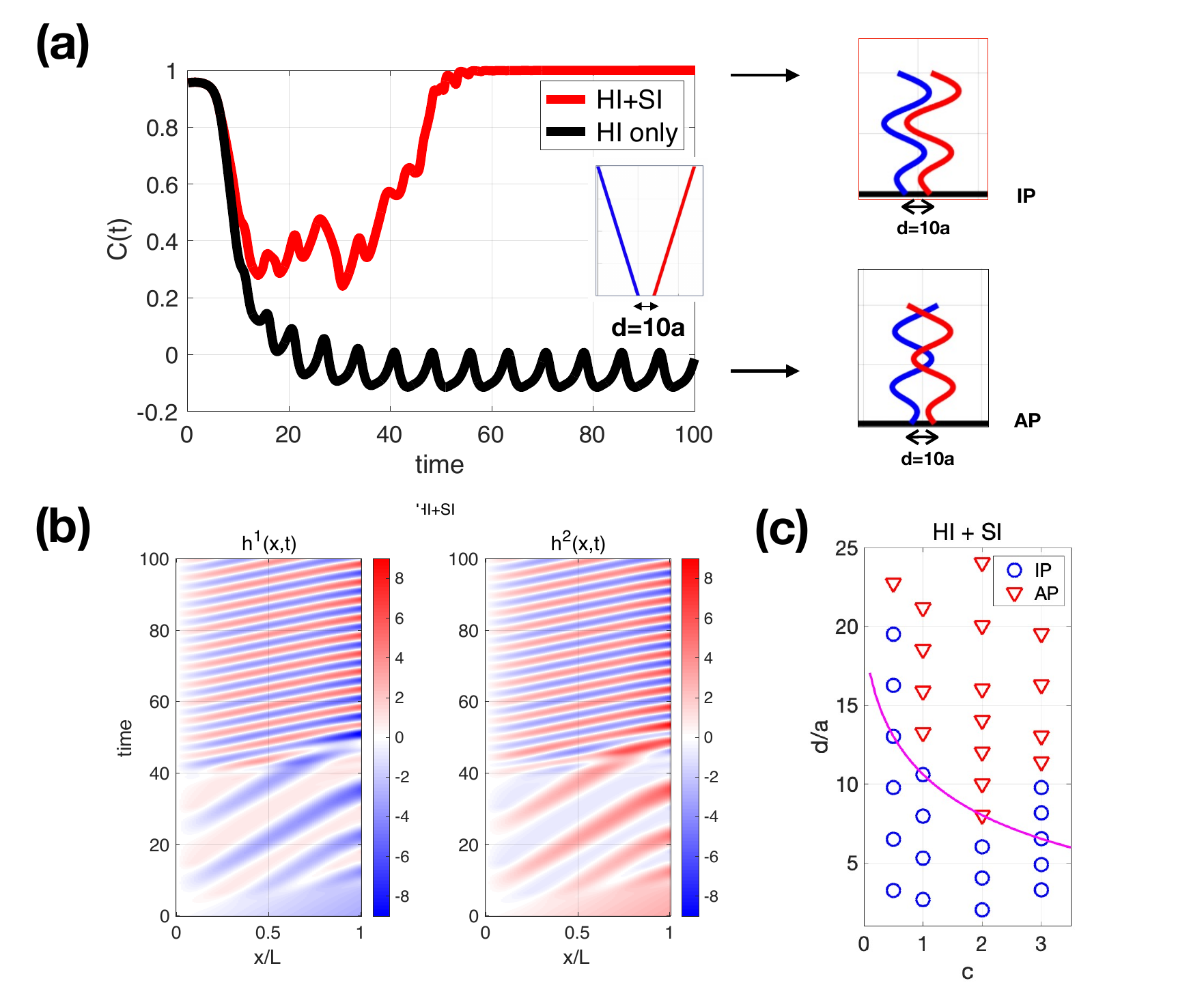}
	\caption{Dynamics of a pair of hydrodynamically-coupled filaments from V-shape initial configuration with and without SI (see SM Movie 4). 
		(a) Evolution of correlation functions ($C(t)$) for the two filaments coupled either only through HIs (black) or through HIs and SIs (red).
		Depicted in the insets are the steady-state configurations in (top) IP and (bottom) AP synchrony. 
		(b) The kymographs of two filaments ($h^1(x,t)$ and $h^2(x,t)$) coupled through HIs and SIs. 
		(c) Phase diagrams of in-phase (IP) and anti-phase (AP) synchronization between two active filaments from the V-shaped initial condition for various values of $c$ and $d$. 
		The steady state at given $c$ and $d$ is determined by using $C(t)$ (Eq.~\ref{eqn:Ct}), such that $C(t)>0.6$ for the IP state, and 
		$C(t)\leq 0.6$ for the AP state. 
		The solid line in magenta plots the $c$-dependent fluctuations of a single filament ($\delta h(c)$, Eq. (\ref{eq:Ac})). 
	}
	\label{fig:APsync}
\end{figure}

In the absence of SIs, depending on the initial condition, 
the two hydrodynamically coupled filaments demonstrate either in-phase (IP) or anti-phase (AP) synchronization after a transient time~\cite{man2020multisynchrony}. 
Adding the short-range SI little affects the hydrodynamically coupled filaments already in IP synchrony; however, the same SI alters the two-filament configuration in AP synchrony into the one in IP synchrony, particularly, when the amplitude of undulation is greater than the inter-filament separation ($\langle \delta h\rangle\gtrsim d$).



%
Two-filament dynamics starting from a V-shape configuration  demonstrates the effect of SI dramatically (see SM Movie 4).
The filaments, coupled only through HI, 
converge to a configuration in AP synchrony (the black line in Fig.~\ref{fig:APsync}(a)).
However, an introduction of SI destabilizes the AP configuration, nudging the system towards the one in IP synchrony (the red graph in Fig.~\ref{fig:APsync}A)~\cite{elfring2009hydrodynamic,leptos2013antiphase,man2020multisynchrony}. 
The kymographs of $h^1(x,t)$ and $h^2(x,t)$ depict the transition from AP ($h^2= -h^1$) to IP ($h^2= h^1$) at $t\gtrsim 40$, accompanied with the decrease of beat period from $(1+\epsilon_h)T_o$ to $(1-\epsilon_h)T_o$ (Fig.~\ref{fig:APsync}(b)). 

The phase diagram of the steady state configurations of two filaments  
(Fig.~\ref{fig:APsync}(c)) shows the stability of AP and IP modes with respect to $c$ and $d$
~\cite{guo2018bistability,goldstein2016elastohydrodynamic}, along with their phase boundary which can be described using the $c$-dependent amplitude of transversal displacement, $\delta h(c)$, obtained 
by assuming that a single traveling wave for the tail part is given as
$h(x,t)=\delta h\cos(k x-\omega t)$. 
Since $h(x,t)$ should satisfy Eq.~1, we obtain the  amplitude of transversal displacement~\cite{goldstein2016elastohydrodynamic}: 
\begin{align}
\langle \delta h\rangle (c)= \frac{2\sqrt{2-\left[\left(c/8\right)^{2/3}+1/3\right]}}{\sqrt{3}\left[\left(c/8\right)^{2/3}+1/3\right]}. 
\label{eq:Ac}
\end{align}
IP state is stable for $c$ satisfying $\langle\delta h\rangle(c)>d$~\cite{elfring2009hydrodynamic,man2020multisynchrony}.
%

\begin{figure}[t]
	\includegraphics[width=\linewidth]{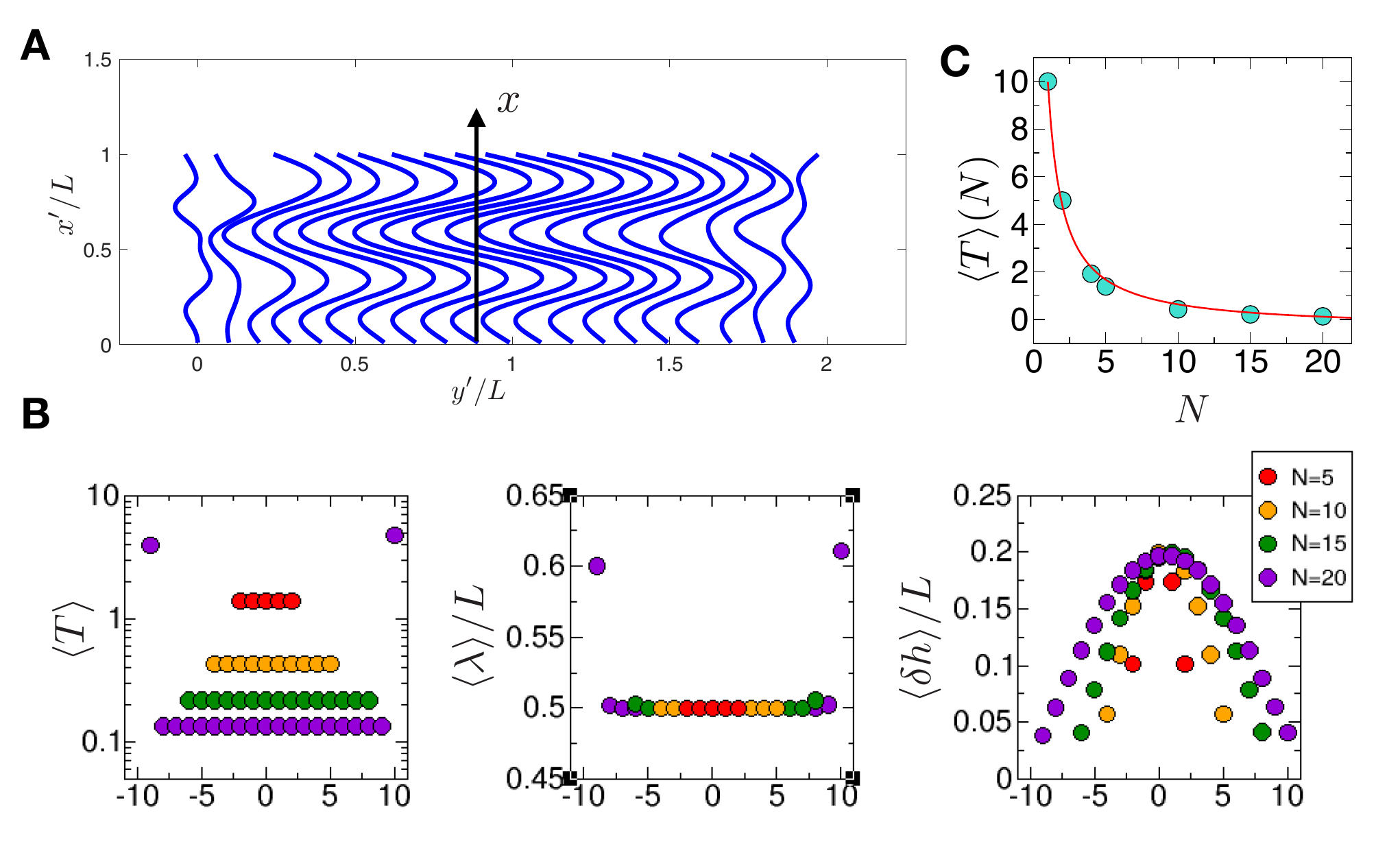}
	\caption{Dynamics among filaments at $\theta=0^\circ$ and $c=1$ with varying $N$.  
		(a) Configuration of $N=20$ filaments in in-phase synchrony. 
		(b) $\langle T\rangle$, $\langle \lambda\rangle/L$, $\langle \delta h\rangle/L$ calculated for $i=1,\ldots, N$ filament. 
		For the demonstration, the filament in the middle is centered. 
				(c) Mean period of filament dynamics in synchrony, $\langle T\rangle$ as a function of $N$. 
		The red line is the fit using Eq.~5. 
	} 
	\label{fig:0_sync}
\end{figure}

\section{Many filaments with non-periodic boundary conditions}
We consider a case where 
a group of active filaments with non-periodic boundary conditions to study the impact of the number of filaments $N$ on the collective dynamics at two tilt angles, $\theta=0^{\circ}$ and $60^{\circ}$.

For a set of $N$-filaments with $\theta=0^{\circ}$ (see Fig.~\ref{fig:0_sync}(a) for a configuration of $N=20$ filaments), the filaments beat in perfect synchrony, so that the mean periods ($\langle T\rangle$) and wavelengths ($\langle \lambda\rangle$) are identical for different filaments except for those at the ends in the case of $N=15$ and $20$, whereas the amplitude of transversal displacement ($\langle \delta h\rangle$) is the largest for the filaments in the middle (Fig.~\ref{fig:0_sync}(b)). 
Notably, the wavelength of the waveform is effectively identical as $\langle\lambda\rangle/L=0.5$, irrespective of $N$; however, 
the mean period ($\langle T\rangle$) decreases with $N$ (Fig.~\ref{fig:0_sync}(c)). Thus, the wave propagation speed ($v_c=\langle\lambda\rangle/\langle T\rangle$) along the centerline ($x$) increases with $N$ (see SM Movie 4).  

\begin{figure}[t]
	\includegraphics[width=\linewidth]{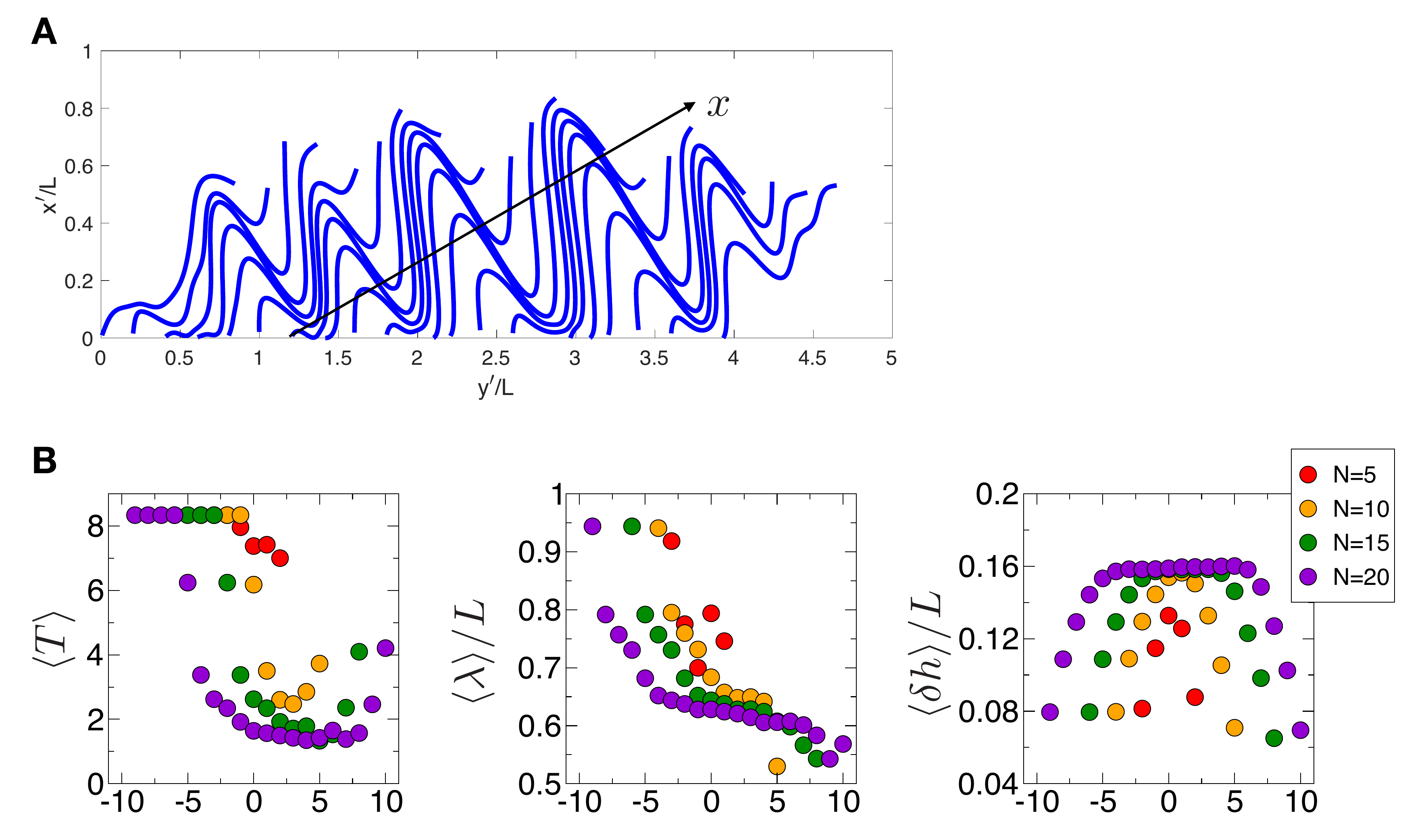}
	\caption{Dynamics among filaments at $\theta=60^\circ$ and $c=1$ with varying $N$. 
	(a) Configuration of N = 20 filaments in synchrony.   
        (b) Same as Fig.~\ref{fig:0_sync}(b). 
	}
	\label{TMW}
\end{figure}

In fact, Eq.~5 with the fit parameters, $T_o\approx 9.98$, $\bar{b}=1$, and $\epsilon^{\rm eff}\approx 0.52$, well describes the periods of collective motion of filaments ($\theta=0^\circ$) for varying $N$.   
Note, however, that in the collective wave generated from the $N$ filaments, there is neither the phase lag between the adjacent filaments nor the significant change in the waveform. 
The waveform of each filament bears the LR symmetry with no wave packet on average being transmitted to left or right along the $y'$-axis (see SM Movie 5).

In contrast, for $\theta=60^{\circ}$, it is clearly seen that MCWs travel from left to right along the $y'$-axis, although the beat dynamics of filaments are only partially synchronized (see SM Movie 6).

For $N=20$, the wave characteristics of the filaments for the two different tilt angles averaged over all the filaments excluding those contributing at the boundary are given as follows:  
(i) $\langle \lambda\rangle=0.5$ $L$ and $\langle T\rangle=0.13$ for $\theta=0^{\circ}$;  
(ii) $\langle \lambda\rangle=0.71$ $L$ and $\langle T\rangle=2.07$ for $\theta=60^{\circ}$. 
Thus, the wave propagation speeds along the centerline 
are  $v_c^{0^\circ}=76.9$ and $v_c^{60^\circ}=6.76$. 
Note that $v_c^{0^\circ}$ is greater than $v_c^{60^\circ}$ by 10 folds. 
The transversal fluctuations, $\langle \delta h\rangle$, are also greater  as well for $\theta=60^\circ$. 
It is of note that more significant reduction is made in $\langle T\rangle$ with increasing $N$ at $\theta=0^\circ$ than at $\theta=60^\circ$, which is primarily due to the enhancement of HI in multiple filaments in full synchrony at $\theta=0^\circ$ in comparison with those in partial synchrony at $\theta=60^\circ$.  

\section{Propulsion Force}
With the resistive force theory \cite{lauga2009hydrodynamics}, we compute the anisotropic drag forces on the $i$th filament in the perpendicular and parallel directions with respect to the centerline of the filament body as follows. 
\begin{align}
	&&f_{\perp}(x^i,t)=-\xi_{\perp} \cos \psi \frac{\partial h^i}{\partial t}  \nonumber \\
	&&f_{\parallel}(x^i,t)=-\xi_{\parallel} \sin \psi \frac{\partial h^i}{\partial t} 
\end{align}
where $\psi$ is the angle formed between the filament orientation at $x^i$ and the centerline ($x$-axis) and $\xi_{\perp}=2\xi_{\parallel}=4\pi\mu/(\ln L /a)$.
Then, the drag forces along the $x$- and $y$-axis are calculated as 
\begin{align}
	f_{x}&= - f_{\perp}\sin \psi  +  f_{\parallel}\cos \psi\nonumber\\
	&=( \xi_{\perp}   -\xi_{\parallel}) \cos \psi \sin \psi\frac{\partial h^i}{\partial t}\nonumber \\
	&\approx \frac{1}{2} \xi_{\perp}\frac{\partial h^i}{\partial x} \frac{\partial h^i}{\partial t} \nonumber\\
	f_{y}&= f_{\perp}\cos \psi + f_{\parallel}\sin \psi\nonumber\\&=-(\xi_{\perp}  \cos^2 \psi + \xi_{\parallel} \sin^2 \psi)  \frac{\partial h^i}{\partial t}\nonumber\\
	&\approx -\xi_{\perp}    \frac{\partial h^i}{\partial t} 
	\label{Eq:dragxy}
\end{align}
where we have used the condition of small undulation in comparison with 
the body length ($\sin \psi \approx \partial_x h^i$, $\cos \psi \approx 1$, and $\sin^2 \psi \approx 0 $) to obtain the approximate expressions for the drag forces. 

Thus, the total drag forces generated by the $i$th filament along the $x$ and $y$-axes are
\begin{align}
F^i_{drag,x}&=\frac{1}{2} \xi_{\perp}\int_0^L \frac{\partial h^i}{\partial x} \frac{\partial h^i}{\partial t}  dx^i\nonumber \\
F^i_{drag,y}&=-\xi_{\perp}  \int_0^L  \frac{\partial h^i}{\partial t}  dx^i
\end{align}
From the force balance, the propulsion force exerted to the fluid due to the beat motion of a filament  
are $F^i_{prop,x}=-F^i_{drag,x}$ and $F^i_{prop,y}=-F^i_{drag,y}$. 
Thus, the total propulsion forces are obtained by summing over the contributions from $N$ filaments, $F_{prop,x}=\sum_{i=1}^{N} F^i_{prop,x}$ and $F_{prop,y}=\sum_{i=1}^{N} F^i_{prop,y}$, and the tilt angle ($\theta$) dependent total propulsion forces along the $x'$- and $y'$-axes are 
\begin{align}
	&&F_{prop,x'}= F_{prop,x}\cos\theta - F_{prop,y}\sin \theta\nonumber \\
	&&F_{prop,y'}=F_{prop,x}\sin\theta + F_{prop,y}\cos \theta
\end{align}
The mean propulsion forces over the period of beat motion in the $x'$- and $y'$-directions are
\begin{align}
\bar{F}_{prop,x'}&=\bar{F}_{prop,x}\cos{\theta}\nonumber\\
\bar{F}_{prop,y'}&=\bar{F}_{prop,x}\sin{\theta}\nonumber\\
\end{align}

Next, the velocity of the fluid flow at position $(x,y)$ at time $t$ generated by the undulatory motion of the $i$th filament is 
\begin{align}
	\bold{u}^i(x,y,t) = - \int_0^L \bold{G}(r)\cdot \bold{f}(x^i,t) dx^i
\end{align}
where $\bold{f}=(f_x,f_y)$ defined in Eq.~(\ref{Eq:dragxy}), $\bold{G}(r)$ is in Eq. (S5), $\bold{r}=(x-x^i,y-h^i)$, and $r=|\bold{r}|$.
The flow velocity along the $x'$ and $y'$-axes are 
\begin{align}
u^i_{x'}&=u^i_x \cos\theta - u^i_y \sin \theta \nonumber \\
u^i_{y'}&=u^i_x \sin \theta + u^i_y \cos\theta. 
\end{align}
Therefore, the fluid flow generated by the collective motion of filaments is $u_{x'}=\sum_i^{N} u^i_{x'}$ and $u_{y'}=\sum_i^{N} u^i_{y'}$. 
Fig.~5 shows the mean values of $u_{y'}$ at location $x'$ from the $y'$-axis.

\section{Viscous dissipation and work production}
First, the mean viscous dissipation per filament (Fig.~5A) is evaluated as: 
\begin{align}
	\langle\dot{Q}\rangle&=-\frac{1}{N}\sum_i^{N}\int_0^L f_y \frac{\partial h^i}{\partial t} dx^i\nonumber\\
	&\approx \frac{1}{N}\sum_i^{N}\int_0^L \xi_{\perp}\left(\frac{\partial h^i}{\partial t}\right)^2 dx^i. 
\end{align}

Next, since the swimming velocity along the $x$- and $y$-directions \cite{lauga2009hydrodynamics,wiggins1998flexive},  resulting from the propulsion  force $\bar{F}_{prop,x}$ of a filament are 
$U_{x}=\bar{F}_{prop,x}/\xi_{\parallel}L$ and $U_{y}=0$, 
the velocities along the $x'$- and $y'$-directions can be expressed as
${U}_{x'}=	{U}_{x}\cos\theta$ and ${U}_{y'}={U}_{x}\sin\theta$. 
Consequently, the rate of work generated and transmitted to the fluid, 
which enhances the fluid flow along the $y'$-direction, is:
\begin{align}
	\langle\dot{W}\rangle=\bar{F}_{drag,y'}\bar{U}_{y'}=\frac{(\bar{F}_{drag,x})^2\sin^2\theta}{\xi_{\parallel}L}
\end{align}

\end{document}